\documentclass[usedcolumn,usenatbib]{mn2e}
\usepackage{graphicx}
\usepackage{times}
\newlength{\colwidth}
\newlength{\fullwidth}
\newlength{\twothirdwidth}
\newlength{\greywidth}
\newlength{\grauwidth}
\newcommand{\mkpc}{\mathrm{~kpc}}

\title[Quasar host galaxy star formation activity from multicolour
data]{Quasar host galaxy star formation activity from\\ multicolour data
 \thanks{Based on observations made at the European Southern Observatory, 
 	La Silla, Chile, and on observations made with the Nordic
	Optical Telescope, La Palma.}
}
\author[K.~Jahnke, B.~Kuhlbrodt, L.~Wisotzki]{K.~Jahnke,$^{1,2}$ B.~Kuhlbrodt,$^{1,2}$ L.~Wisotzki$^{1,3}$ \\
	$^1$Astrophysikalisches Institut Potsdam, An der Sternwarte
16, 14482 Potsdam, Germany\\
        $^2$Hamburger Sternwarte, Gojenbergsweg 112, 21029 Hamburg, Germany\\
	$^3$Universit\"at Potsdam, Am Neuen Palais 10, 14469 Potsdam, Germany}
\date{00 00 00}
\pagerange{\pageref{firstpage}--\pageref{lastpage}}
\pubyear{200?}

\begin{document}

\maketitle

\label{firstpage}

\begin{abstract}
We investigate multicolour imaging data of a complete sample of 19 low
redshift ($z<0.2$) quasar host galaxies. The sample was imaged in four
optical (\textsl{BVRi}) and three near-infrared bands
(\textsl{JHK}$_s$). Galaxy types, structural parameters and robust
host galaxy luminosities are extracted for all bands by means of
two-di\-men\-sio\-nal deblending of galaxy and nucleus. For the disc
dominated fraction of host galaxies (Sa and later) the optical and
optical-to-NIR colours agree well with the average colours of inactive
galaxies of same type. The bulge dominated galaxies (E/S0) on the
other hand appear a significant 0.35~mag bluer in $(V-K)$ than their
inactive counterparts, being as blue as the discs in the sample. This
trend is confirmed by fitting population synthesis models to the
extracted broad band SEDs: the stellar population age of the bulge
dominated hosts lies around a few Gyr, much younger than expected for
old evolved ellipticals. Comparison to other studies suggests a strong
trend for stellar age in elliptical host galaxies with
luminosity. Intermediately luminous elliptical hosts have comparably
young populations, either intrinsically or from an enhanced star
formation rate potentially due to interaction, the most luminous and
massive ellipticals on the contrary show old populations. The
correspondence between the nuclear activity and the blue colours
suggests a connection between galaxy interaction, induced star
formation and the triggering of nuclear activity. However, the
existence of very symmetric and undisturbed disks and elliptical host
galaxies emphasised that other mechanisms like minor merging or gas
accretion must exist.
\end{abstract}


\section{Introduction}
Since the discovery of massive black holes in the centres of normal
galaxies and an identical relation between bulge and black hole mass
for active and inactive, quasar host galaxies start to be accepted as
not-so-different members of the galaxy population than previously
thought. The question remains how different quasar hosts are from
`normal' galaxies. Is a quasar phase part of every host galaxies life?
The connection of the masses of galactic bulges and central black
holes suggests a common evolutionary path. Nuclear activity might thus
be a special phase of black hole fuelling. Obvious candidates for
triggering active galactic nuclei (AGN) include major and minor merger
in the scenario of hierarchical clustering, as well as internal
instabilities. The effects of these events in the shape of nuclear
activity and, potentially, enhanced induced star formation thus might
give an opportunity to study a generic phase in the evolutionary cycle
of galaxies.

Single optical or near infrared (NIR) bands are generally sufficient
to characterise morphological properties like galaxy types, host and
nuclear luminosities, or to conduct environment studies
\citep[e.g.][]{mcle95b,mclu99}. It was also attempted to detect and
quantify events of merging or interaction from pure morphology of the
visible host galaxy \citep[e.g.][]{hutc92}.

To assess the dominant stellar populations in the host, spectral
information is needed. While one optical colour already permits a
characterisation of young stellar populations, the restframe NIR on
the other hand holds information about the old stars, dominating the
host in mass.

Ideally, spectra of the host galaxy would be needed, spanning all the
optical and NIR domains. However, due to the unavoidable contamination
of the final host galaxy spectrum with a nuclear component, increasing
with decreasing angular resolution, spectroscopy is so far of limited
use for luminous quasars (QSOs), already at low redshifts.

With this study we attempt to assemble spectral information on quasar
hosts by combining optical and NIR broad band data \citep[as first
performed by][]{koti94}. At the price of a coarse sampling of the SED
compared to spectroscopy, the deblending of nuclear and stellar
components in broad band images is by now a robust and reliable
technique.

In the present paper we present multicolour data for a complete sample
of low redshift quasars. The sample selection and observations are
described in Section~2. The deblending into nuclear and host
contributions is presented in Section~3, followed by a description of
the photometry (Sect.~4). In Section~5 we discuss the results,
including notes on individual objects. The derived host galaxy colours
are compared to those of inactive galaxies in Section~6, while
Section~7 contains results from simple stellar population model
fits. In Sections~8 and 9 we present an overall discussion and some
conclusions.

Throughout this article we use
a cosmology with $H_0=50$~km~s$^{-1}$~Mpc$^{-1}$, $q_0=0.5$ and
$\Lambda=0$.

\section{Sample and observations}

\begin{table*}
\caption{\label{t_sample}
Objects in the samples. Shown are the redshift $z$, the total apparent
$V$ magnitude and the total $K$-corrected $V$-band magnitude of the
object. The last column shows the radio classification from the VLA in
comparison to $B$-band fluxes, `RL' for radio-loud, `RQ' for
radio-quiet, and `?'  for unknown radio properties. `RQ?'  marks four
low luminosity objects without radio measurements.
}
\begin{tabular}{ll...c}
Object & Alternate\
name&\multicolumn{1}{c}{$z$}&\multicolumn{1}{c}{$V$}&\multicolumn{1}{c}{$M_V$}&Radio properties\\
\hline
HE\,0952--1552&  		& 0.108 & 15.8&-23.3&RQ\\
HE\,1019--1414&  		& 0.077 & 16.1&-22.3&RI\\
HE\,1020--1022& PKS\,1020--103 	& 0.197 & 16.6&-23.9&RL\\
HE\,1029--1401&  		& 0.085 & 13.7&-24.9&RQ\\
HE\,1043--1346&  		& 0.068 & 15.7&-22.5&RQ?\\
HE\,1110--1910&  		& 0.111 & 16.0&-23.2&RQ\\
HE\,1201--2409&  		& 0.137 & 16.3&-23.5&RQ\\
HE\,1228--1637&  		& 0.102 & 15.8&-23.2&RQ\\
HE\,1237--2252&  		& 0.096 & 15.9&-23.0&RQ\\
HE\,1239--2426&  		& 0.082 & 15.6&-23.0&RQ\\
HE\,1254--0934&  		& 0.139 & 14.9&-24.7&RI\\
HE\,1300--1325& R\,12.01 	& 0.047 & 14.9&-22.5&RQ?\\
HE\,1310--1051& PG\,1310--1051 	& 0.034 & 14.9&-21.7&RQ?\\
HE\,1315--1028&  		& 0.099 & 16.8&-22.1&RQ\\
HE\,1335--0847&  		& 0.080 & 16.3&-22.3&RQ\\
HE\,1338--1423& R\,14.01 	& 0.041 & 13.7&-23.3&RQ?\\
HE\,1405--1545&  		& 0.194 & 16.2&-24.2&RQ\\
HE\,1416--1256& PG\,1416--1256 	& 0.129 & 16.4&-23.2&RI\\
HE\,1434--1600&  		& 0.144 & 15.7&-24.1&RL\\
\end{tabular}
\end{table*}

\begin{figure}
\includegraphics[bb = 35 56 313 314,clip,angle=0,width=\colwidth]{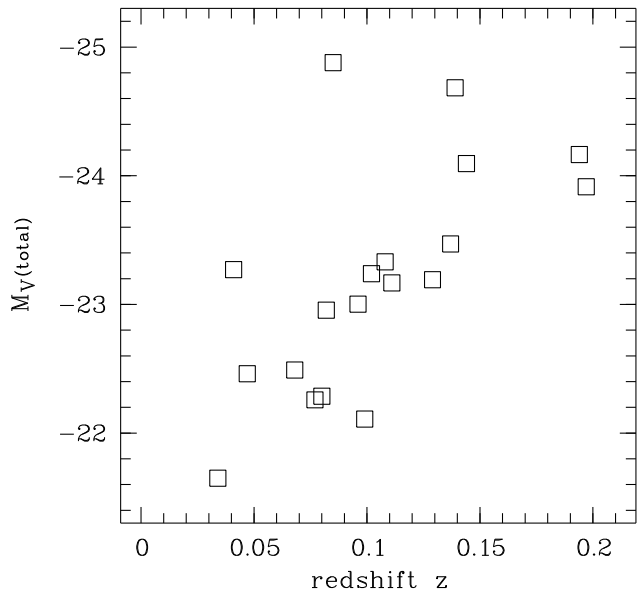}
\caption{\label{z_v} 
Redshifts and total absolute {\it V}-band magnitudes of the quasar
sample.
}
\end{figure}

We compiled a sample of 19 objects with $z<0.2$, drawn from the
Hamburg/ESO survey \citep[HES,][]{reim96,wiso96,wiso00}. The sample is
statistically complete, comprising all quasars above a well-defined
flux limit from a sky area of 611$\;$deg$^2$. It is a low-$z$ subset
of the sample defined by \citet{koeh97} to study the luminosity
function of quasar nuclei. Its distribution in redshifts and absolute
magnitudes is shown in Fig.~\ref{z_v}. The sample represents QSOs of
moderate optical luminosities when compared to the total population at
all redshifts. The radio properties of the sample have recently been
determined with the VLA (Gopal-Krishna et al.\ in preparation). We
classified the objects along the definition by \citet{kell89}, using
the ratio $R \equiv F_\mathrm{5\,GHz}/F_B$ of (nuclear)
radio-to-optical fluxes, and the three classes of radio-quiet ($R<1$),
radio-intermediate ($1<R<10$) and radio-loud objects ($R>10$). For the
four lowest redshift objects ($z<0.07$) no radio information is
available, but since the sample is optically selected, they are likely
to be radio quiet. Two objects in the sample are radio-loud, three are
radio-intermediate, the rest is radio quiet. Table~\ref{t_sample}
lists the objects, redshifts, apparent and absolute $V$ magnitude and
the radio properties.

In order to evenly sample the spectral energy distribution (SED) from
the optical to the NIR, we imaged the 19 objects of the sample in
seven filters: Johnson {\it B}, Bessel {\it V} and {\it R}, Gunn {\it
i} in the optical, and {\it J}, {\it H}, and $K_s$ in the NIR (from
now on referred to as \textit{BVRIJHK} for simplicity). The three NIR
bands are primarily sensitive to the old stellar populations in the
host galaxy, while the optical bands are trace the continuum of young
stars. As a bonus, this even sampling of the SED will allow to compare
the properties of this sample to those at higher redshifts without the
need for $K$-corrections.

\begin{table}
\caption{\label{t_obsprop}
Telescope and observation properties. Given
are band, telescope and instrument used, number of objects,
integration times, and pixel scale.
}
\begin{tabular}{llcll}
Band& Instrument& $N_\mathrm{obj}$ &$t_\mathrm{int}$ [s]& Scale\\
\hline
\textsl{B} &ESO 3.6m/EFOSC2 &12 & 30& 0\farcs61\\
\textsl{B} &ESO 3.6m/EFOSC1 &3  & 30& 0\farcs34\\
\textsl{B} &ESO 1.54m/DFOSC &2  &  380/730& 0\farcs39\\
\textsl{B} &NOT/ALFOSC      &2  & 1000/1200& 0\farcs19\\
\textsl{V},\textsl{R},\textsl{i} &ESO 1.54m/DFOSC &19 &300--1200& 0\farcs39\\
\textsl{J},\textsl{H},\textsl{K}$_s$ &ESO NTT/Sofi &19 &160--900& 0\farcs29
\end{tabular}
\end{table}

For the majority of the observations ESO telescopes were used. The
optical data were mainly obtained with DFOSC at the 1.54m Danish
telescope, supplemented by $B$-band images from
\citet{jahn03e}. Two additional frames were obtained with ALFOSC at
the 2.5m Nordic Optical Telescope (NOT) on La Palma. All NIR data were
taken with the ESO NTT and SOFI. Table~\ref{t_obsprop} gives a summary
of the telescopes used, as well as typical integration times and
instrumental pixel scales.

The images in the $V$, $R$, $I$ and NIR bandpasses reach similar
surface brightness levels for most objects. This is different for the
EFOSC1/2 $B$-band images, which were originally taken as acquisition
frames for spectroscopy. With their integration time of only 30\,s
they are much shallower than the rest of the data and were exchanged
for deeper images from other campaigns in four cases.

\subsection{Data reduction}
For data reduction in the optical $V$, $R$, and $I$-bands we created
flatfield frames from the object frames themselves (`superflats'). The
object fields are sufficiently empty and contained enough sky
background for this task, yielding higher quality flatfields than
normal twilight sky flats which were also taken. For the ALFOSC and
EFOSC $B$-band images twilight and dome flats were used,
respectively. Bias frames were obtained nightly in all
cases. Otherwise a standard reduction procedure was followed.

The NTT/SOFI NIR data suffered from a detector peculiarity, a time and
illumination dependent bias pattern. Theoretically, the spatially
variable bias in NIR arrays is -- under standard conditions --
automatically removed when subtracting a sky frame from the data
frame, thus the bias is spatially variable but the pattern stable in
time. On the other hand, the SOFI chip/controller unit used produced a
bias pattern that depended on the time since last reset and on the
illumination level, thus on the amount of sky background. Since at the
time of reduction the effect was not finally investigated by ESO, we
developed a special reduction procedure that was able to remove these
patterns with in most cases negligible residual structures.

For all objects and bands -- except EFOSC1/2 $B$-band data and the
optical bands of HE\,1434--1600 -- more than one image per band was
observed with the individual frames dithered by some tens of
arcseconds between integrations. Variance weighted stacking plus
rejection of outlier pixel values for these dithered frames allowed to
remove artefacts created by cosmic ray hits, dead columns, hot pixels
and other localised errors. The variance weighted stacked images were
then further processed by multi-component deblending.

\section{Quasar multi-component deblending}\label{sec:mc_qsomodelling}
The critical step in the image analysis is a proper separation of
nuclear and host galaxy components. We created a package for
two-dimensional multi-component deblending of quasars by fitting a
parametrised host model plus a nuclear contribution. The developement
and testing of the algorithms are described in detail by
\citet*{kuhl02}. For the present sample we chose to use both two and
three component models for the QSO, consisting of the nucleus,
represented by the PSF, and the host galaxy, represented by one or two
analytical two-dimensional functions, convolved with the PSF.

\subsection{Model fits}\label{sec:bb_modelling}
For the PSF model we used an elliptical Moffat function, to allow for
non-circular symmetry. This was necessary to model the optical
distortions of the focal reducer-type instruments EFOSC, DFOSC and
ALFOSC. Several stars in each field were used to define the PSF.

The PSF model fit yields the best estimate for the shape of the PSF,
thus the shape of the QSO \emph{nucleus}. In the following second step
the amount of flux contained in the nucleus is fitted simultaneously
with the shape and flux of the host galaxy component. The latter is
represented by an analytic galaxy model consisting of either a
de~Vaucouleurs spheroidal \citep{deva48}, an exponential disc
\citep{free70}, or a combination of both:

\begin{eqnarray}   
 F_{\mathrm{sph}}(r) &=& F_{\mathrm{sph},0} \exp\left[-7.67\,\left(\frac{r}{r_{1/2}}\right)^{1/4}\right]
 \label{eqn:sph}\\   
 F_{\mathrm{disc}}(r) &=& F_{\mathrm{disc},0} \exp\left(-1.68\,\frac{r}{r_{1/2}}\right)
 \label{eqn:disc}   
\end{eqnarray}   
where $F_0$ is the central surface brightness and the `radius' $r$ is
a function of $x$ and $y$
\begin{equation}
    \label{eqn:rad}
    r^2 \:=\: \frac{1-\epsilon(2-\epsilon)\cos^2(\alpha-\phi)}{(1-\epsilon)^2}\left(x^2+y^2\right)\:,
\end{equation}
with $\tan\alpha = y/x$, $\epsilon$ the ellipticity and $\phi$ the
position angle of the isophotes. In this form the scale length
$r_{1/2}$ is always the radius of the isophote containing half of the
total flux.

The galaxy models are convolved numerically with the PSF to
incorporate the effect of the seeing also for the host. As a result of
the deblending fit we receive models of nucleus and host component(s).

We restricted the pixels used for deblending to a region of interest
defined as an elliptical anullus centred on the nucleus. This was
chosen to be large enough to contain all visible flux of the object
($>$99 per cent) or, as in the case of HE\,1043--1346 (see
Fig.~\ref{fig:bb_grey1}), was restricted to the central region where
the host has a well defined structure, excluding the spiral arms
further out. Geometrical or physical companion objects as well as
strong asymmetric components like tidal tails were masked out by
modifying the variance frame. These areas were excluded from the fit,
to satisfy the model assumptions, i.e.\ a smooth galaxy
component.

\subsection{Homogeneous treatment for all bands}
In order to guarantee for a consistent procedure in all bands of the
multicolour sample, identical modelling areas and identical masks were
used for all observations of a given object. Only unique artefacts
like cosmic ray hits and dead pixels were added to the masks of
individual frames. The sky background was adjusted to be zero in an
annulus around the object in each frame, determined from
curve-of-growth analysis. Thus background variations within the frame
were accounted for.

One important fact for the deblending is the strongly rising sky
brightness from {\it B} to {\it K}-band. This yields a different
relative weight or S/N of low count (host outskirts) and high count
(central) areas of the quasar when using formal variances. This will
have an influence on the deblending: Even if identical physical
surface brightness distributions were present in, e.g., the $B$ and
$K$-bands, the outskirts of the host will have a larger weight in the
deblending fit in $B$ than in $K$, and vice versa for the nucleus.

However, simulations with different weighting schemes other than using
formal variances resulted in no convincing recipe for weights, and
thus we stayed with formal variances. Nevertheless, this aspect has to
be borne in mind when comparing optical and NIR data of an object.

In order to minimise the influence of this effect on the deblending
process in the different bands, we decided to use two consecutive
quasar deblending iterations. In a first iteration we left all model
parameters free. We then determined global values of the scale lengths
$r_{1/2}$, ellipticities $\epsilon$ and position angles $\varphi$,
valid for all bands. In the second iteration these parameters were
fixed for the deblending, with only the fluxes of nucleus and host
components as free parameters, thus forcing identical host
geometry. We discuss the validity and implications of this approach in
section \ref{sec:mcvarscale}.

\section{Photometry}\label{sec:mcphot}
The multi-component deblending procedure yields models for host
galaxies as well as the nuclei. We decided whether a given host galaxy
is disc or spheroid dominated on the basis of the results of the model
fits, and by visual inspection of the residual images and radial
profiles after model subtraction. The model then gave the overall
morphological parameters of the host.

For photometry of the host the model of the nucleus was used to
subtract the nuclear light contribution from the composite quasar
image, resulting in a host galaxy frame containing all asymmetries,
spiral arms, etc. This method has an advantage over simply integrating
the galaxy model as has been done in the past
\citep[e.g.][]{tayl96,scha00}. Asymmetries like arms or tidal
distortions can be accounted for in this way. Also deblending errors
for the galaxy, e.g.\ in scale length, might have a negligible effect
on the nuclear flux, but can be magnified when the model is directly
integrated. Especially in cases where the type of the galaxy cannot be
firmly established this becomes be a problem. Fits with a spheroidal
model will yield, when integrated, 0.5--1.5\,mag systematically higher
fluxes than the best fitting exponential disc model \citep{abra92}.

\begin{figure}
\includegraphics[bb = 72 559 365 772,clip,angle=0,width=\colwidth]{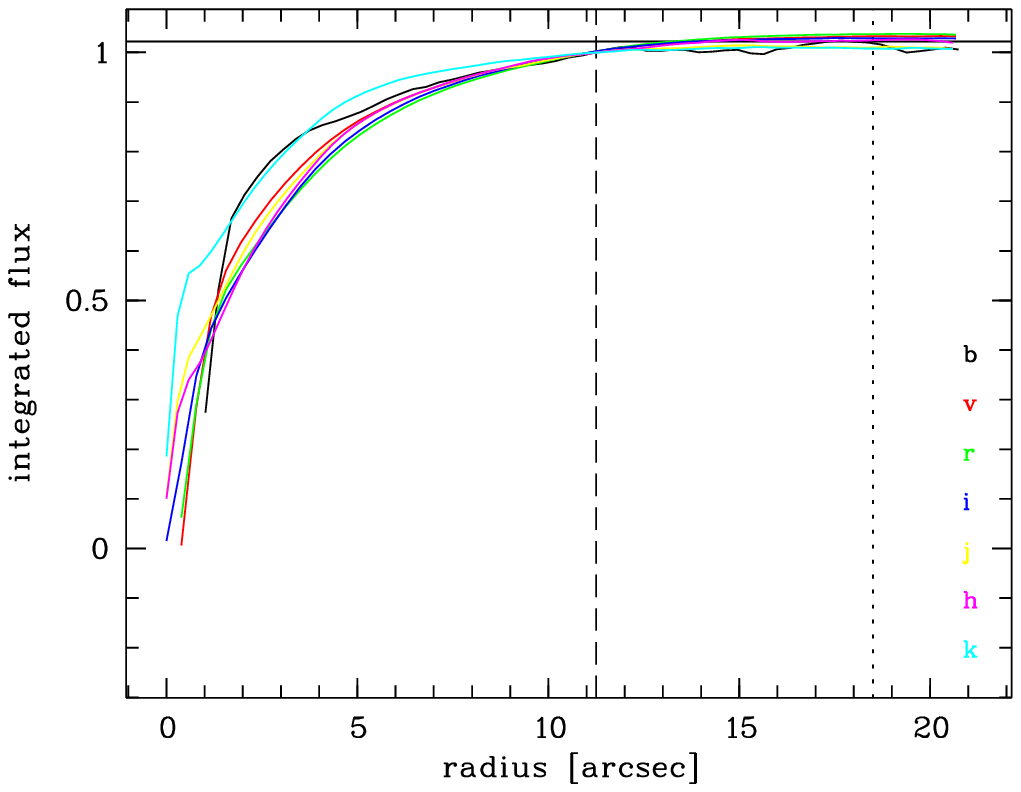}\\
\includegraphics[bb = 72 542 365 772,clip,angle=0,width=\colwidth]{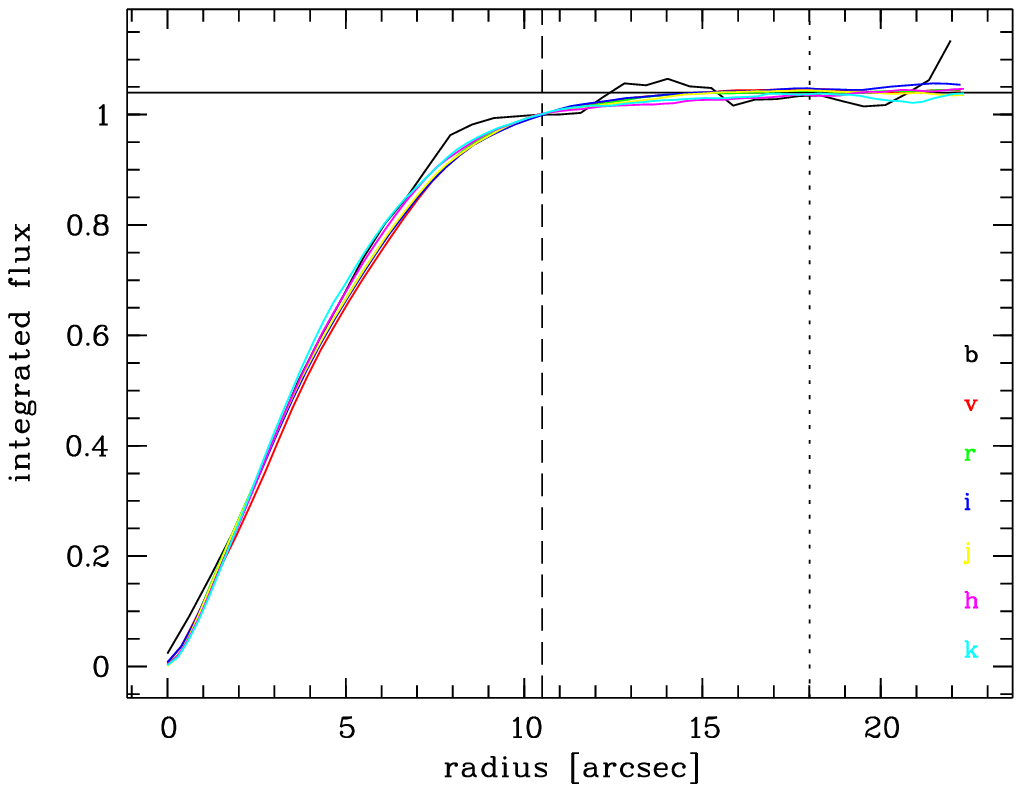}
\caption{\label{fig:photometry_method}
Simultaneous photometry in all bands: Shown are, for two objects, the
curves of growth for all bands. They are normalised to the flux at the
dashed vertical line ($r_\mathrm{eq}$). From the bands with the
highest S/N, in both case all bands except $B$, the flux at the dotted
vertical line ($r_\mathrm{conv}$) is taken as 100 per cent. The
integrated flux for each band at $r_\mathrm{eq}$ is then scaled
accordingly and used as the total flux for this band. This reduces the
influence of background noise at large radii, without integrating to
different isophote levels for different bands.
}
\end{figure}

Photometry of the host image was done via aperture photometry. The
aperture used was an ellipse with the ellipticity and position angle
of the modelled host, and thus fixed for all frames of a given
object. The size of the aperture was chosen to minimise errors from
background noise. Since the observations in the different bands are,
in general, of different depths, curves of growth of the host flux
will begin to be dominated by background noise at different radii for
different bands. We adopted an aperture correction procedure by
defining two radii for photometry (cf.\
Fig.~\ref{fig:photometry_method}): One radius, $r_\mathrm{eq}$,
typically containing 90--98 per cent of the total flux, was chosen such that
the curves of growth in all bands would not deviate much from each
other at larger radii, if the curves of growth were normalised at this
radius. A second radius, $r_\mathrm{conv}$, was chosen at which the
curve of growth of the band(s) with the highest S/N would
converge. The flux in this band inside $r_\mathrm{conv}$ was taken as
100 per cent of the host flux, and the fraction of flux inside was
$r_\mathrm{eq}$ determined. The fraction of flux contained inside
$r_\mathrm{eq}$ for this band was then assumed to be valid for all
bands, and the total flux computed from the value inside
$r_\mathrm{eq}$. This again assumes identical profile shapes and scale
lengths in all bands (see Sect.~\ref{sec:mcvarscale}) but has the
advantage of a homogenous treatment independent of S/N.

Photometric calibration was performed using photometric standard stars
observed during the runs. The {\it B}-band data taken with
EFOSC/EFOSC2 was calibrated using the original photometry from
\citet{koeh97}, with updated galactic dust extinction values. We
used the values from \citet{schl98}, and applied a correction in all
bands. We also applied airmass corrections in the $B$ and $V$-band
using average La Silla extinction coefficients.

\begin{table}
 \caption{\label{t_kcorr}
Adopted $K$-corrections for nucleus and host, in general valid only
for $z<0.2$. The nuclear $K$-corrections are derived from the averaged
QSO SED by \citet{elvi94}. For the host values are taken from
\citet{fuku95} in the optical and \citet{mann01} in the NIR.
}
\begin{center}
\begin{tabular}{lrrr}
Band $X$& \multicolumn{3}{c}{$K(X)$ [mag/$z$]}\\
& \multicolumn{1}{c}{nucleus} & \multicolumn{1}{c}{spheroid} & \multicolumn{1}{c}{disc (Sb)}\\
\hline
$B$&&5.0&3.25\\
$V$&--0.99&2.25&1.0\\
$R$&&1.0&0.75\\
$I$&&0.75&0.5\\
$J$&&0.25&0.15\\
$H$&--1.25&--0.25&--0.25\\
$K$&&--2.4&--2.4\\
\hline
\end{tabular}
\end{center}
\end{table}

Conversion to absolute magnitudes required {\it K}-correction terms,
which we applied separately to quasar nuclei and hosts. For the
nucleus we used the average quasar SED as published by
\citet{elvi94}. For $z<0.5$, a linear relation $K(z) \propto z$ is a
good description, for all relevant bands. For the host we used the
{\it K}-correction terms published by \citet{fuku95} in the optical
and \citet{mann01} in the NIR. We adopted separate values for
predominantly spheroidal and disc hosts, for the latter we assumed
intermediate type Sb discs. In {\it H} and {\it K} the differences
between E and Sb galaxies are negligible. We again approximate the
{\it K}-correction as a linear function of $z$, which yields errors
less than 0.05~mag in all bands for $z<0.2$. The adopted values are
compiled in Table~\ref{t_kcorr}.

\section{Deblending results and analysis}\label{sec:mcresult}
We clearly resolve all host galaxies in all bands.
Figure~\ref{fig:bb_grey1} shows, for each object, a grey scale plot of
the host galaxy after subtraction of the nuclear model, as well as the
radial azimuthally averaged profiles of the different components.

\begin{figure*}
\begin{minipage}[b][\greywidth][t]{4.7cm}
\includegraphics[bb = 95 613 314 777,clip,angle=0,height=\greywidth]{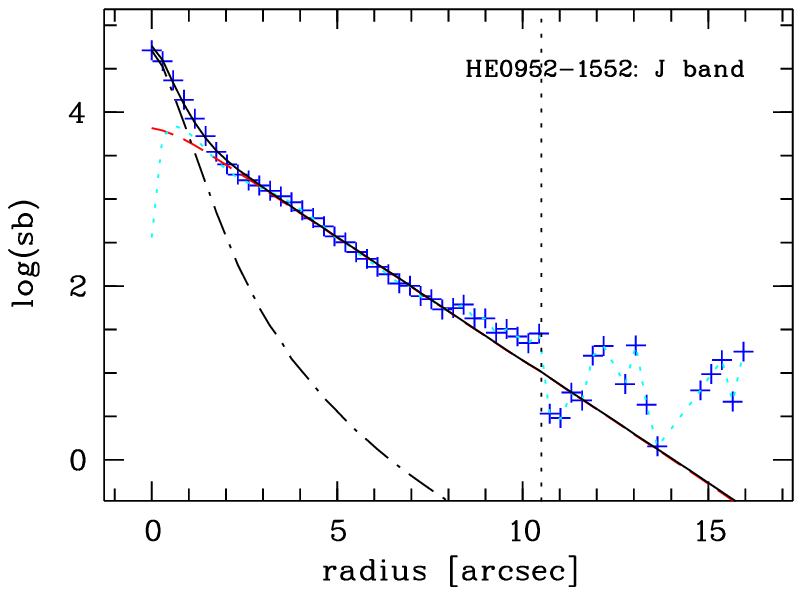}
\end{minipage}
\begin{minipage}[b][3.35cm][t]{\grauwidth}
\includegraphics[bb = 99 99 302 302,clip,angle=0,height=\grauwidth]{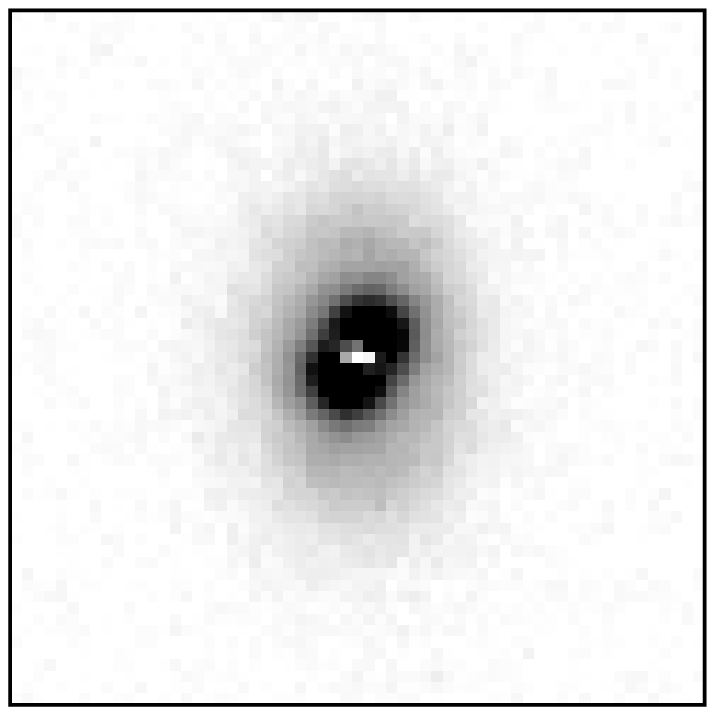}
\end{minipage}
\hfill
\begin{minipage}[b][\greywidth][t]{4.7cm}
\includegraphics[bb = 95 613 314 777,clip,angle=0,height=\greywidth]{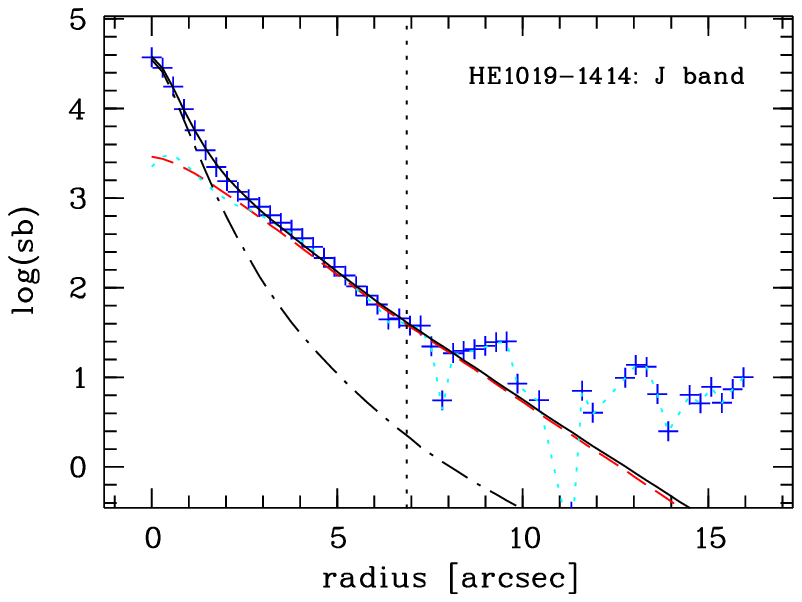}
\end{minipage}
\begin{minipage}[b][3.35cm][t]{\grauwidth}
\includegraphics[bb = 99 99 302 302,clip,angle=0,height=\grauwidth]{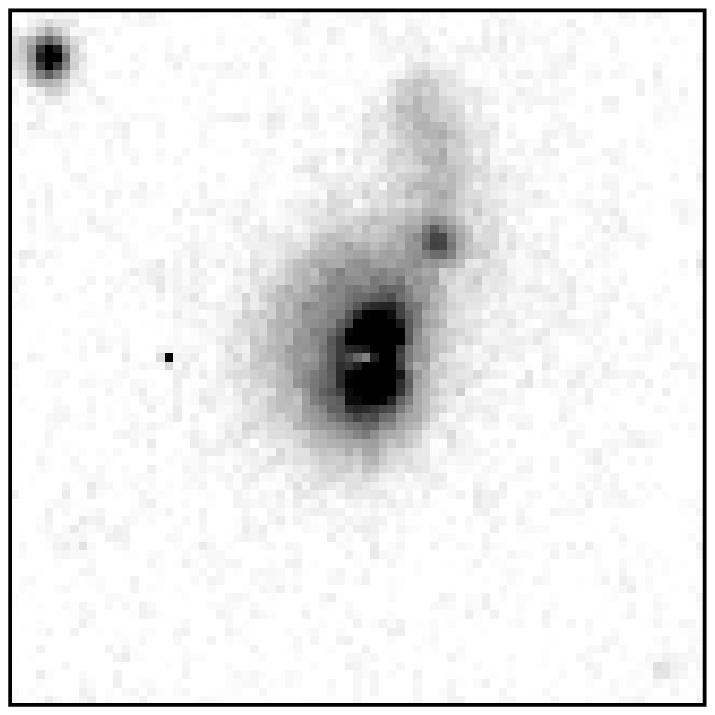}
\end{minipage}\\

\begin{minipage}[b][\greywidth][t]{4.7cm}
\includegraphics[bb = 95 613 314 777,clip,angle=0,height=\greywidth]{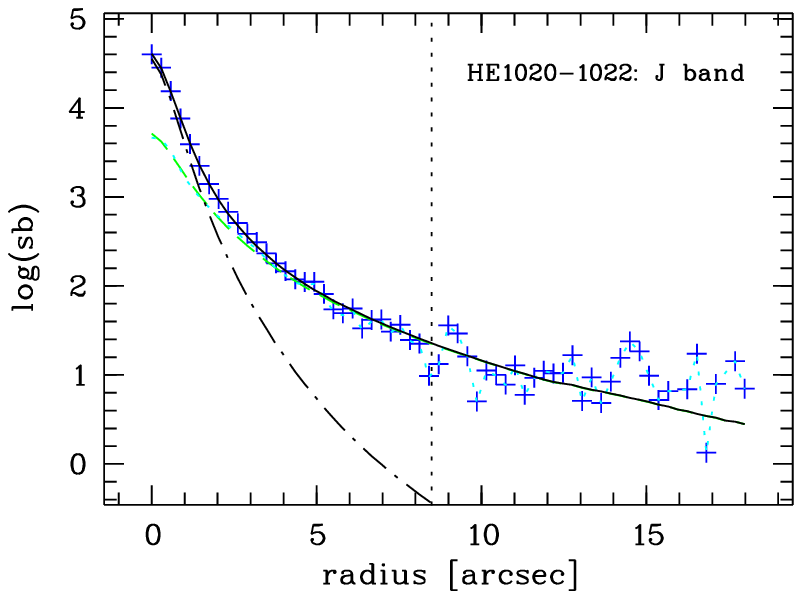}
\end{minipage}
\begin{minipage}[b][3.35cm][t]{\grauwidth}
\includegraphics[bb = 99 99 302 302,clip,angle=0,height=\grauwidth]{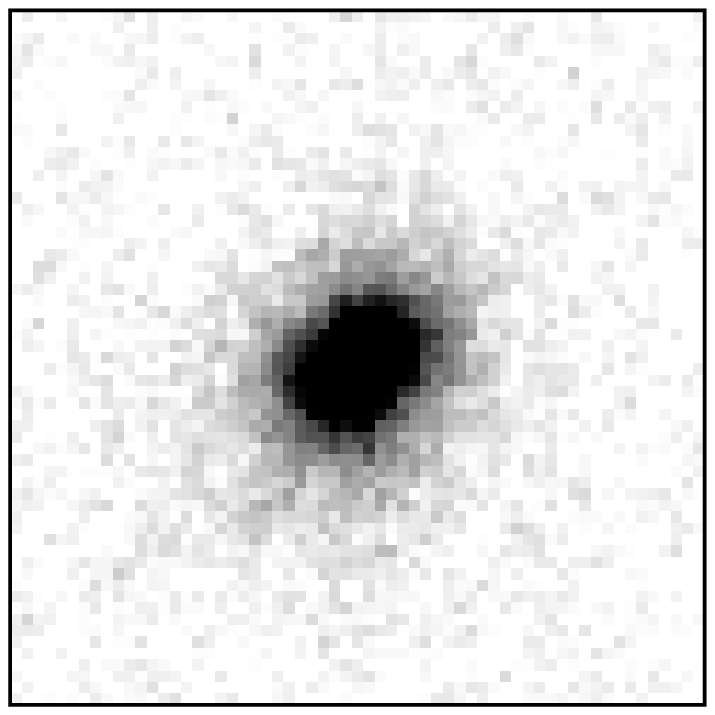}
\end{minipage}
\hfill
\begin{minipage}[b][\greywidth][t]{4.7cm}
\includegraphics[bb = 95 613 314 777,clip,angle=0,height=\greywidth]{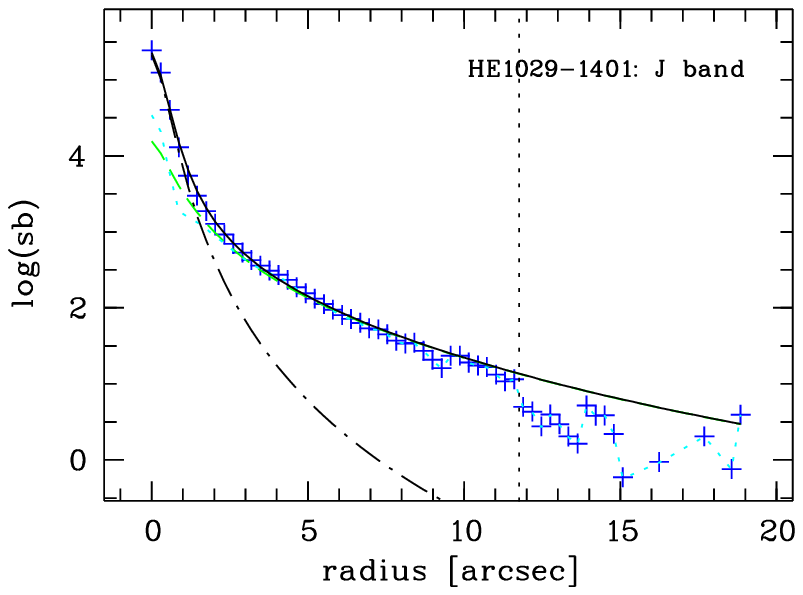}
\end{minipage}
\begin{minipage}[b][3.35cm][t]{\grauwidth}
\includegraphics[bb = 99 99 302 302,clip,angle=0,height=\grauwidth]{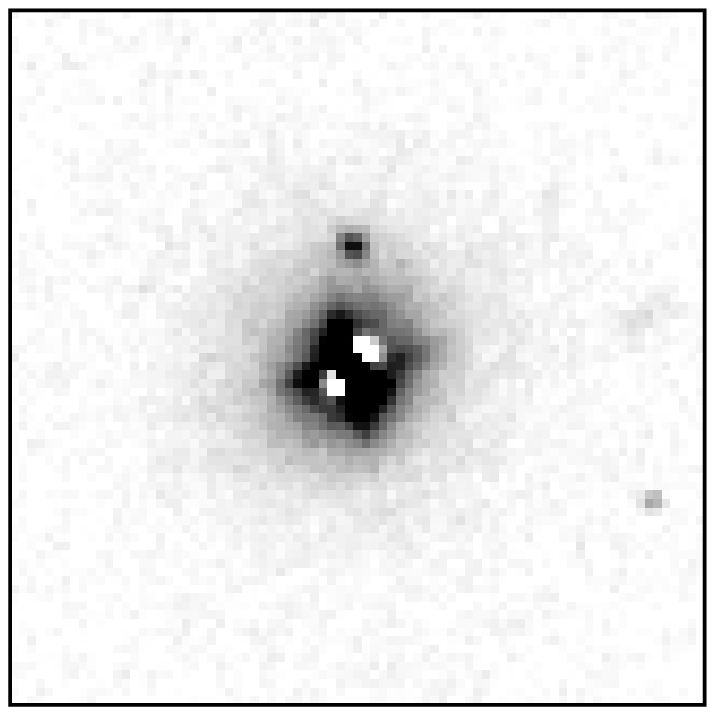}
\end{minipage}\\
\begin{minipage}[b][\greywidth][t]{4.7cm}
\includegraphics[bb = 95 613 314 777,clip,angle=0,height=\greywidth]{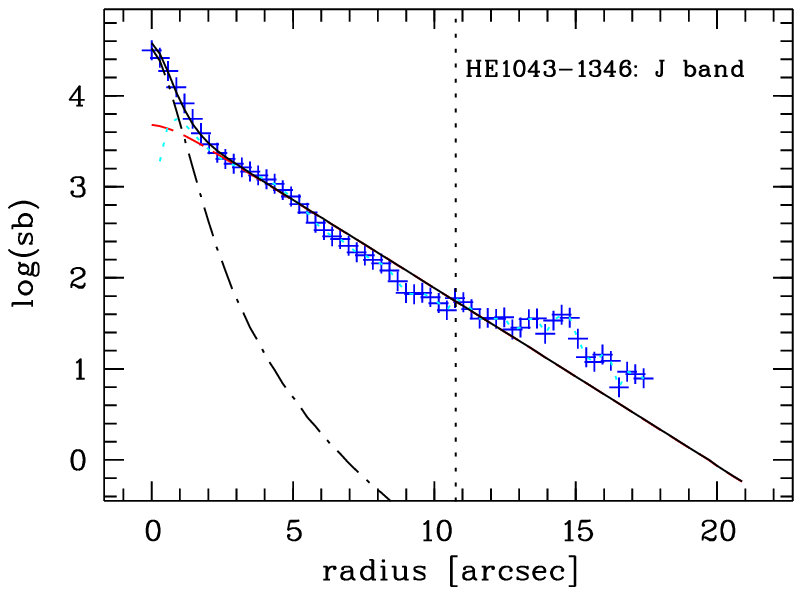}
\end{minipage}
\begin{minipage}[b][3.35cm][t]{\grauwidth}
\includegraphics[bb = 99 99 302 302,clip,angle=0,height=\grauwidth]{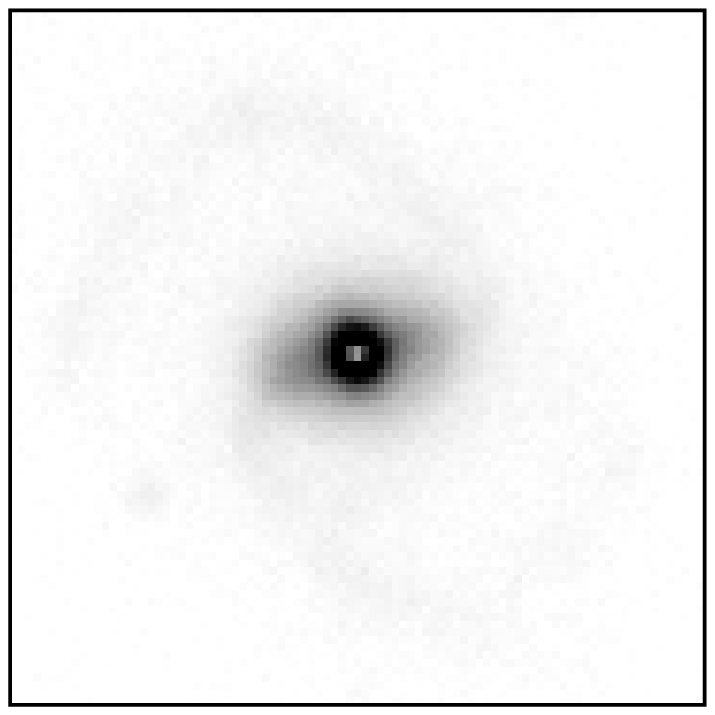}
\end{minipage}
\hfill
\begin{minipage}[b][\greywidth][t]{4.7cm}
\includegraphics[bb = 95 613 314 777,clip,angle=0,height=\greywidth]{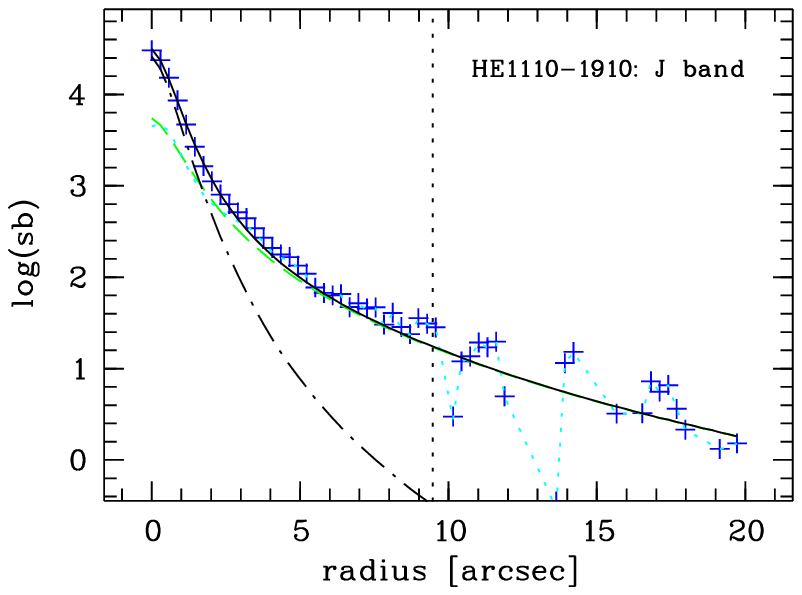}
\end{minipage}
\begin{minipage}[b][3.35cm][t]{\grauwidth}
\includegraphics[bb = 99 99 302 302,clip,angle=0,height=\grauwidth]{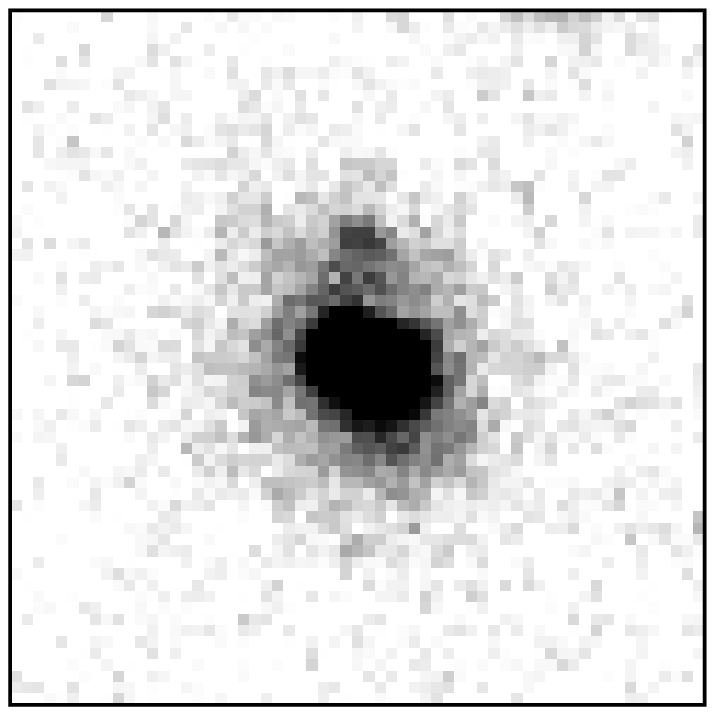}
\end{minipage}\\
\begin{minipage}[b][\greywidth][t]{4.7cm}
\includegraphics[bb = 95 613 314 777,clip,angle=0,height=\greywidth]{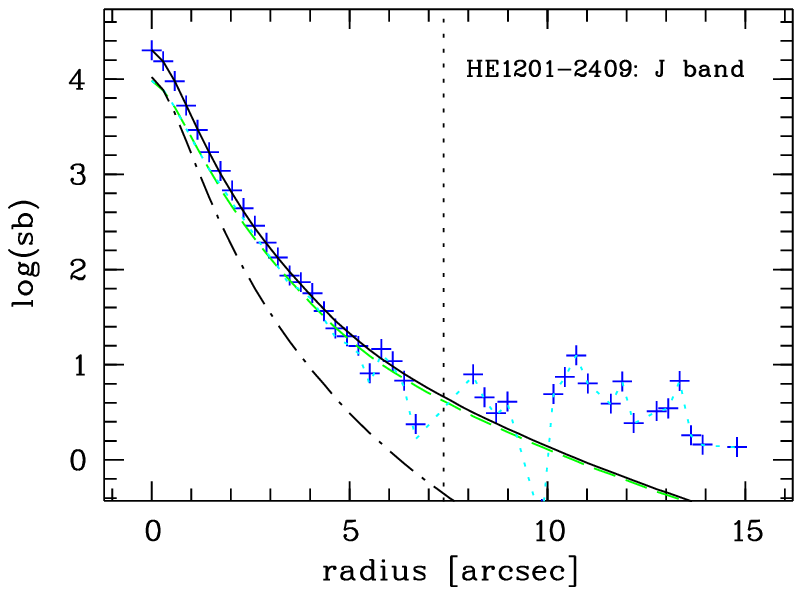}
\end{minipage}
\begin{minipage}[b][3.35cm][t]{\grauwidth}
\includegraphics[bb = 99 99 302 302,clip,angle=0,height=\grauwidth]{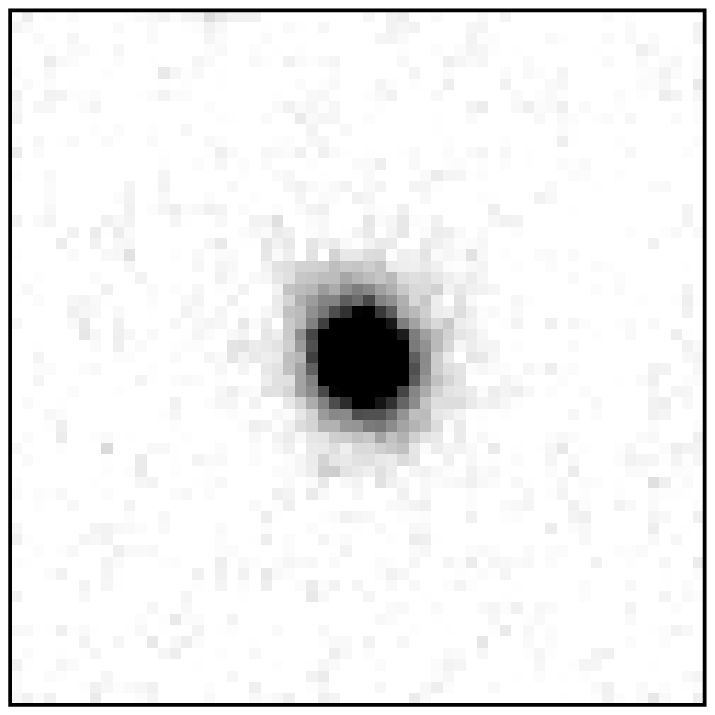}
\end{minipage}
\hfill
\begin{minipage}[b][\greywidth][t]{4.7cm}
\includegraphics[bb = 95 613 314 777,clip,angle=0,height=\greywidth]{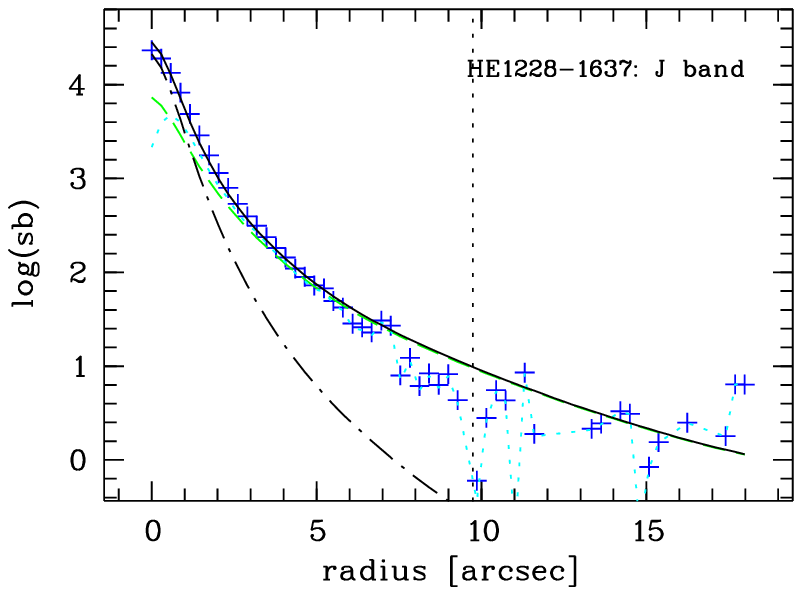}
\end{minipage}
\begin{minipage}[b][3.35cm][t]{\grauwidth}
\includegraphics[bb = 99 99 302 302,clip,angle=0,height=\grauwidth]{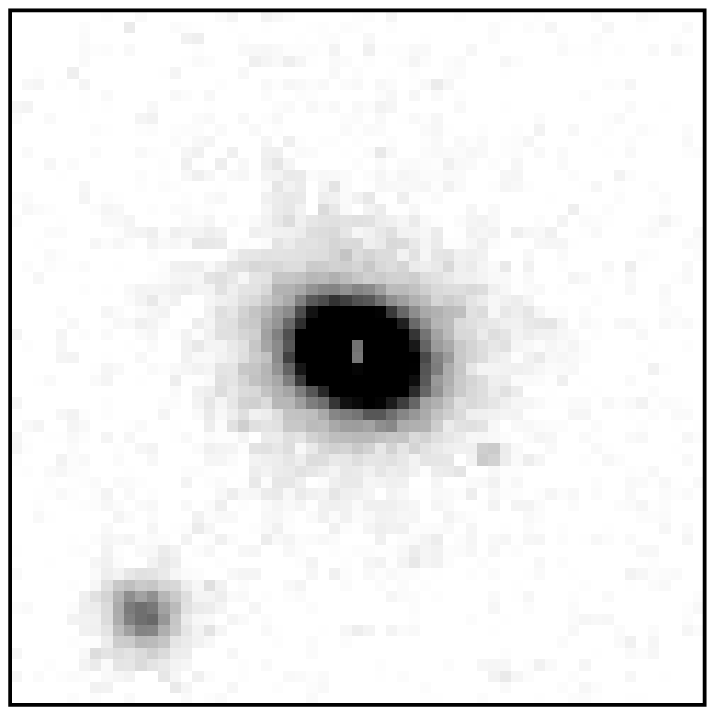}
\end{minipage}\\
\begin{minipage}[b][\greywidth][t]{4.7cm}
\includegraphics[bb = 95 613 314 777,clip,angle=0,height=\greywidth]{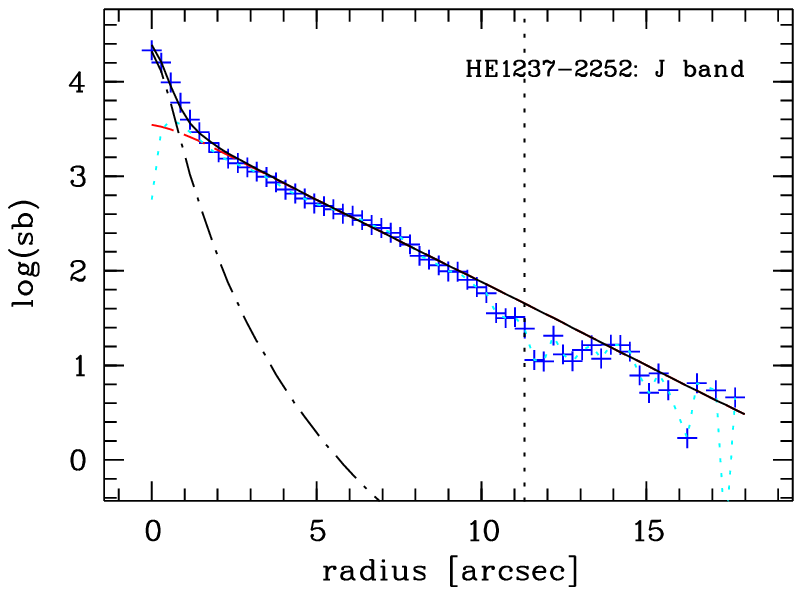}
\end{minipage}
\begin{minipage}[b][3.35cm][t]{\grauwidth}
\includegraphics[bb = 99 99 302 302,clip,angle=0,height=\grauwidth]{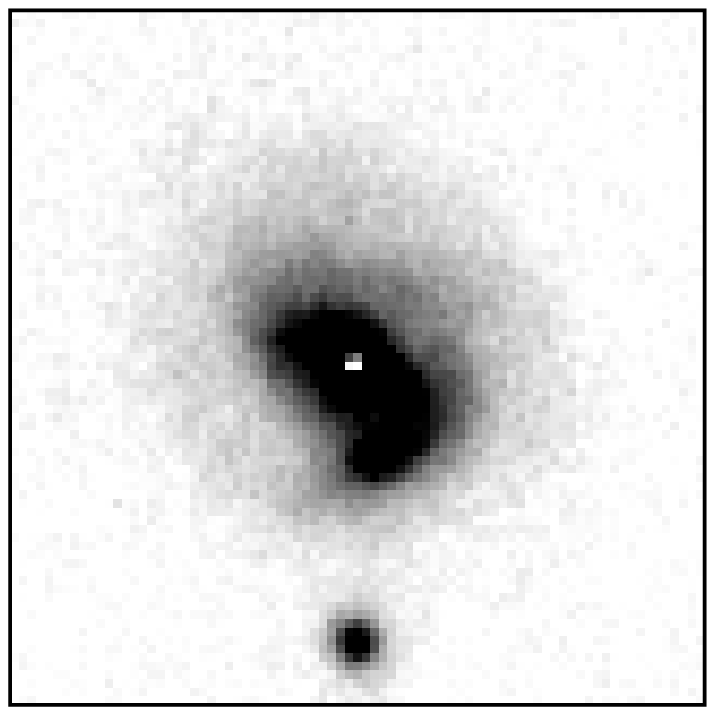}
\end{minipage}
\hfill
\begin{minipage}[b][\greywidth][t]{4.7cm}
\includegraphics[bb = 95 613 314 777,clip,angle=0,height=\greywidth]{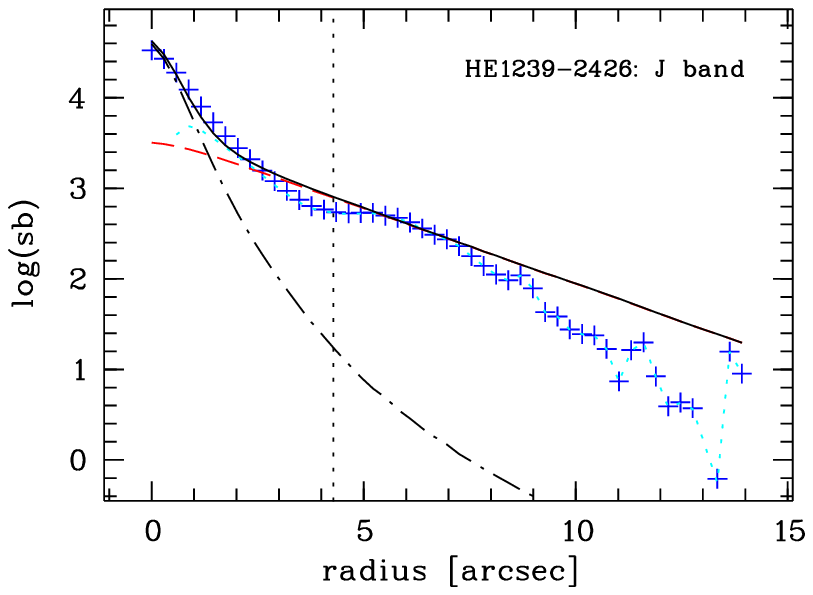}
\end{minipage}
\begin{minipage}[b][3.35cm][t]{\grauwidth}
\includegraphics[bb = 99 99 302 302,clip,angle=0,height=\grauwidth]{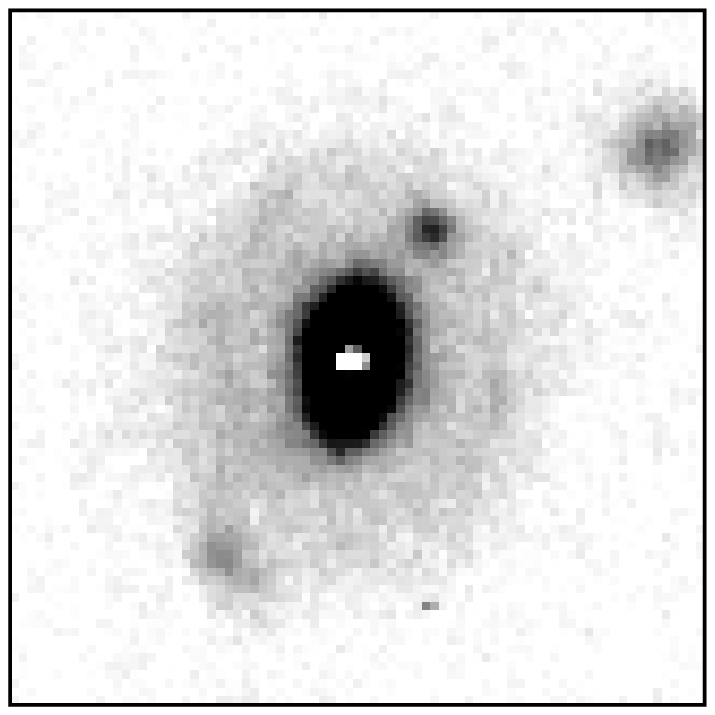}
\end{minipage}\\
\vspace{1cm}
\caption{\label{fig:bb_grey1}
The objects in the multicolour sample: The diagrams show radial
surface brightness profiles of the objects in the $J$-band. Flux is
logarithmic in arbitrary units, radius is in arcseconds. The symbols
mark the data points; the different lines are: nuclear model
(dot-dashed black), host model (long dashed, with green colour for
spheroidals and red colour for discs), combined nuclear plus host
model (solid black), remaining host after subtraction of the nuclear
model (dotted light blue). The vertical line marks the radius of the
ellipse inside which was fitted. The grey scale images show the host
galaxy after subtraction of the nucleus in the NIR $J$-band ($R$-band
for HE\,1043--1346, HE\,1254-0934, and HE\,1310--1051). Images are
centered on the nucleus. The side lengths of the images differ:
$43.5''$ for HE\,1310--1051 and HE\,1338--1423, $35''$ for
HE\,1043--1346 and HE\,1254--0934, $26''$ HE\,1029--1401,
HE\,1237--2252, HE\,1239--2426 and HE\,1300--1325, $23''$ for
HE\,1019--1414 and HE\,1335--0847, and $20''$ for the remaining
objects.
}
\end{figure*}

\begin{figure*}
\begin{minipage}[b][\greywidth][t]{4.7cm}
\includegraphics[bb = 95 613 314 777,clip,angle=0,height=\greywidth]{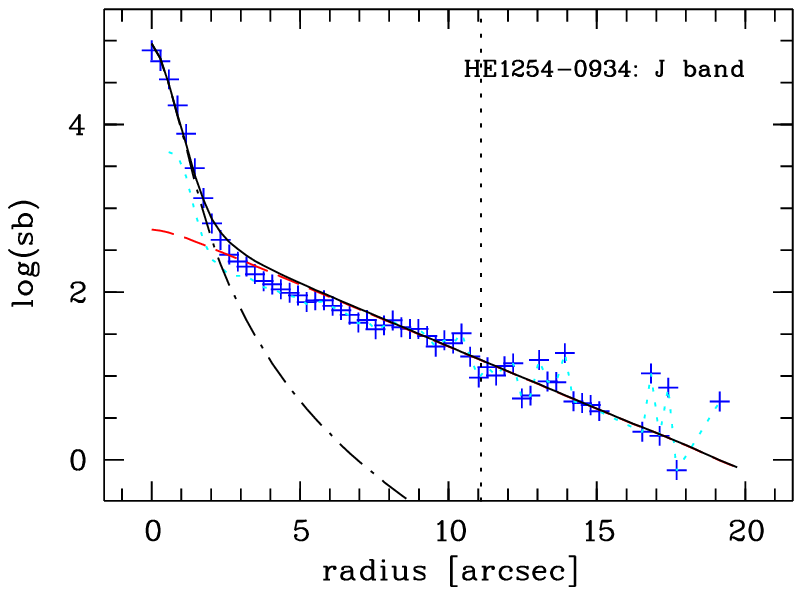}
\end{minipage}
\begin{minipage}[b][3.35cm][t]{\grauwidth}
\includegraphics[bb = 99 99 302 302,clip,angle=0,height=\grauwidth]{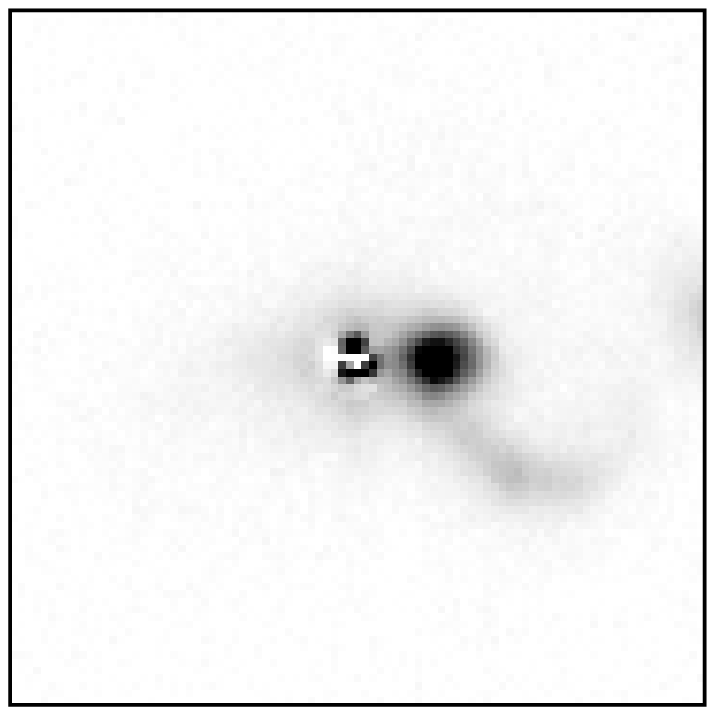}
\end{minipage}
\hfill
\begin{minipage}[b][\greywidth][t]{4.7cm}
\includegraphics[bb = 95 613 314 777,clip,angle=0,height=\greywidth]{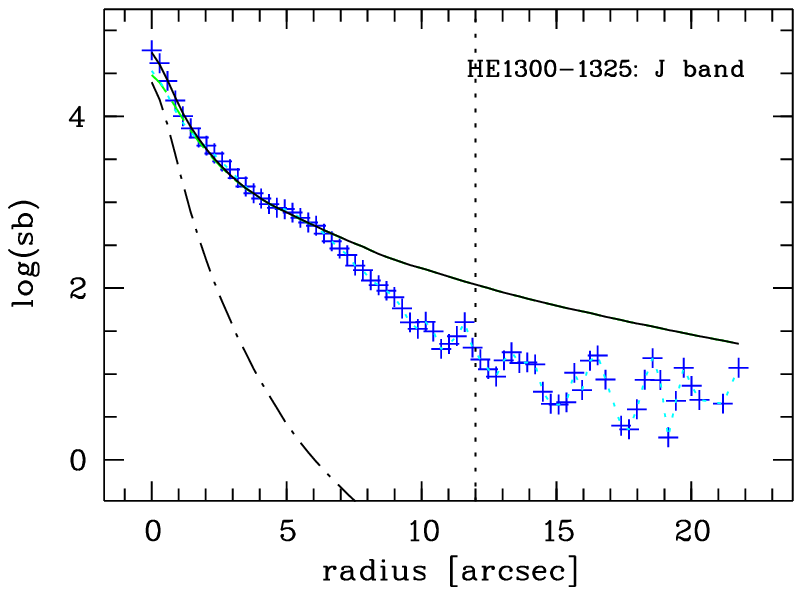}
\end{minipage}
\begin{minipage}[b][3.35cm][t]{\grauwidth}
\includegraphics[bb = 99 99 302 302,clip,angle=0,height=\grauwidth]{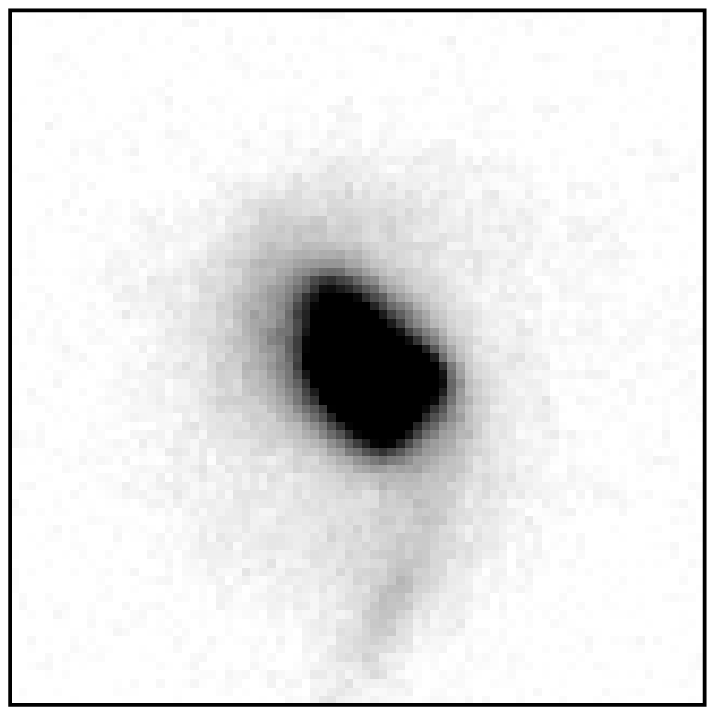}
\end{minipage}\\
\begin{minipage}[b][\greywidth][t]{4.7cm}
\includegraphics[bb = 95 613 314 777,clip,angle=0,height=\greywidth]{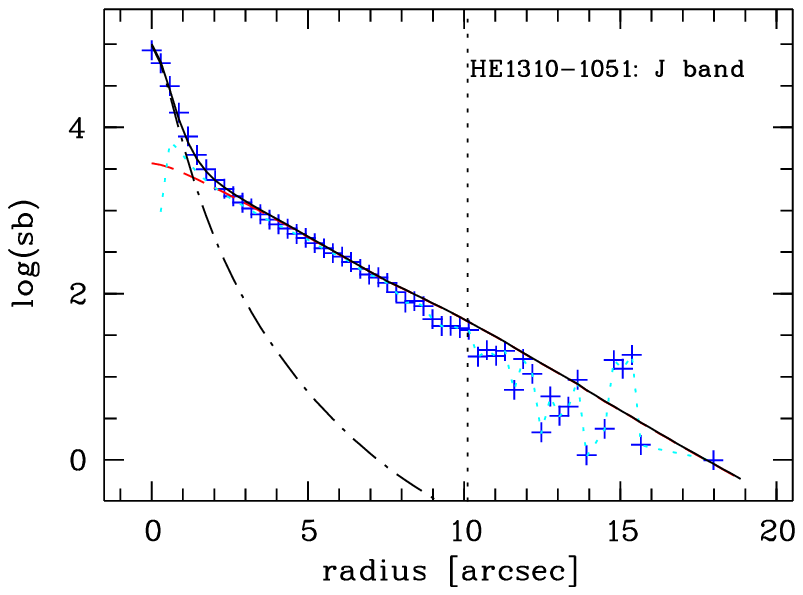}
\end{minipage}
\begin{minipage}[b][3.35cm][t]{\grauwidth}
\includegraphics[bb = 99 99 302 302,clip,angle=0,height=\grauwidth]{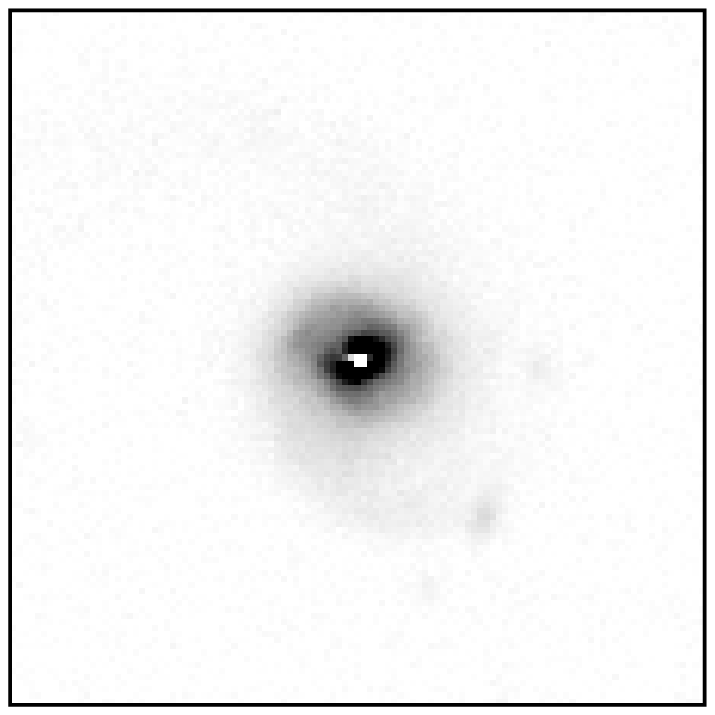}
\end{minipage}
\hfill
\begin{minipage}[b][\greywidth][t]{4.7cm}
\includegraphics[bb = 95 613 314 777,clip,angle=0,height=\greywidth]{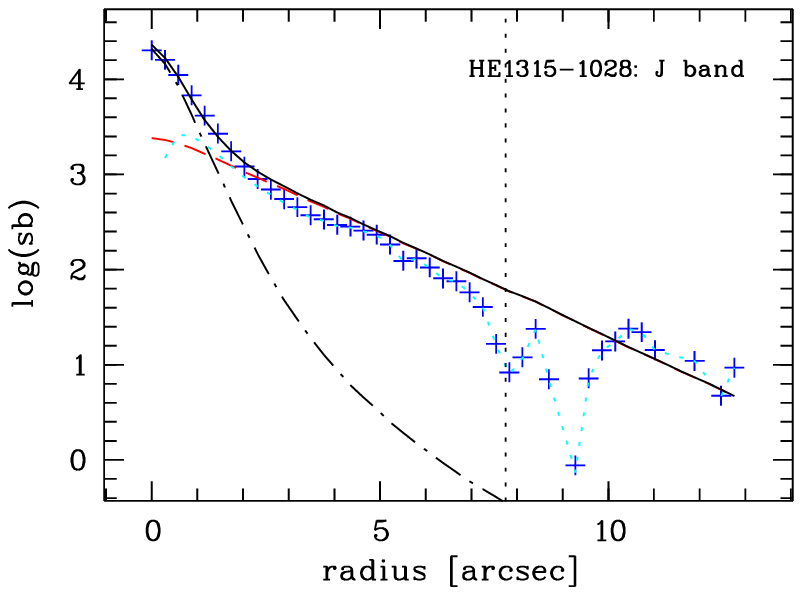}
\end{minipage}
\begin{minipage}[b][3.35cm][t]{\grauwidth}
\includegraphics[bb = 99 99 302 302,clip,angle=0,height=\grauwidth]{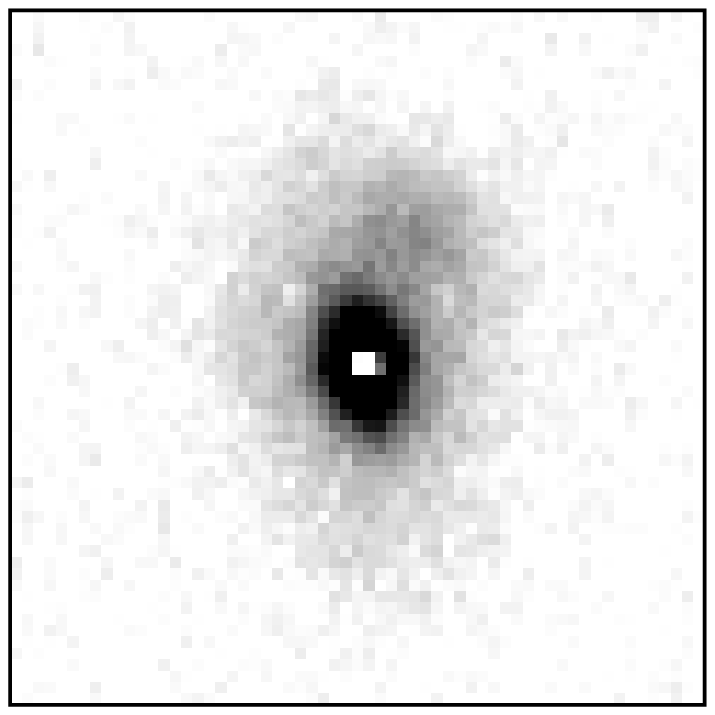}
\end{minipage}\\
\begin{minipage}[b][\greywidth][t]{4.7cm}
\includegraphics[bb = 95 613 314 777,clip,angle=0,height=\greywidth]{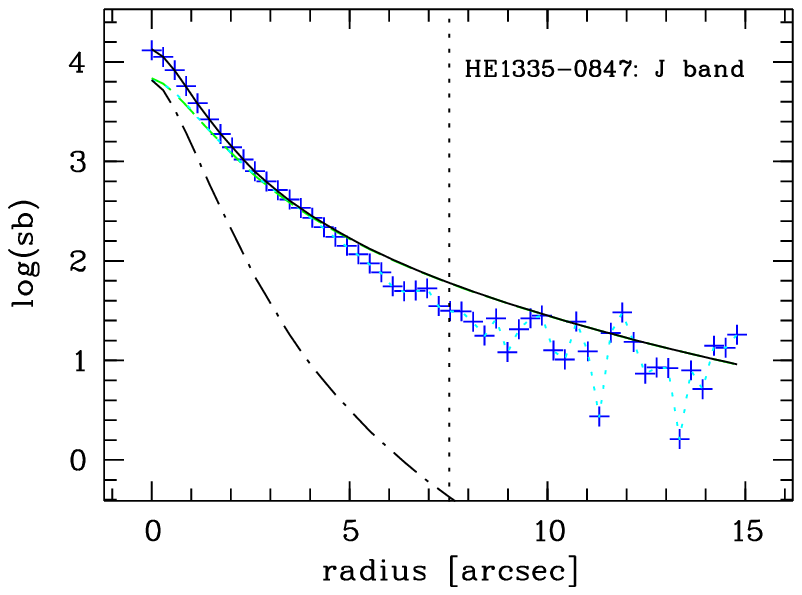}
\end{minipage}
\begin{minipage}[b][3.35cm][t]{\grauwidth}
\includegraphics[bb = 99 99 302 302,clip,angle=0,height=\grauwidth]{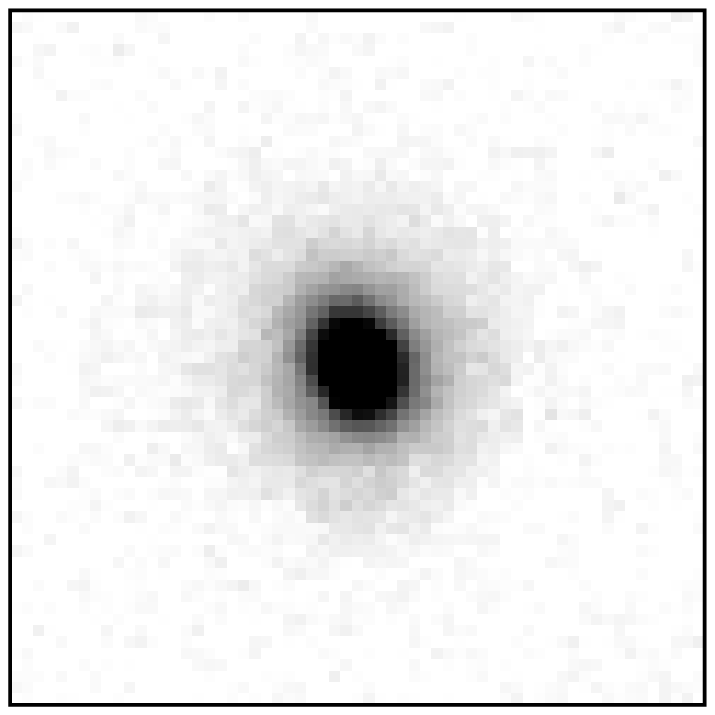}
\end{minipage}
\hfill
\begin{minipage}[b][\greywidth][t]{4.7cm}
\includegraphics[bb = 95 613 314 777,clip,angle=0,height=\greywidth]{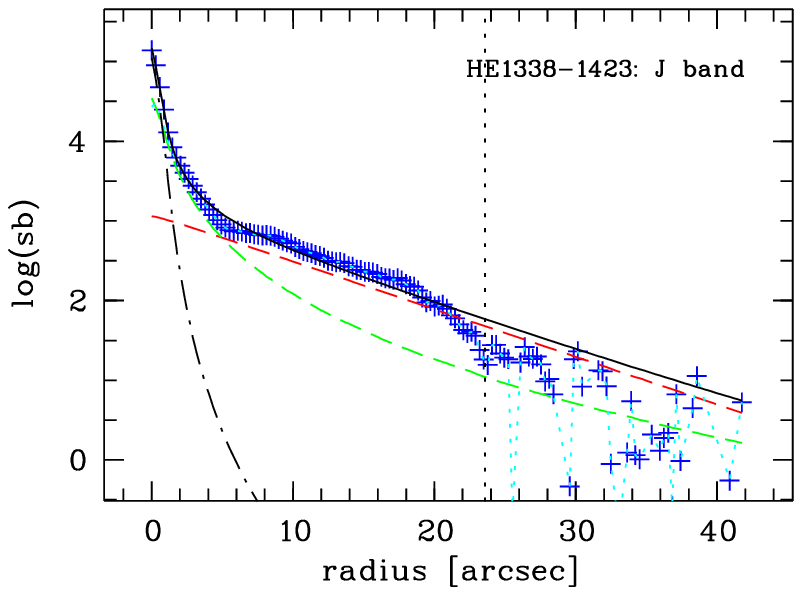}
\end{minipage}
\begin{minipage}[b][3.35cm][t]{\grauwidth}
\includegraphics[bb = 99 99 302 302,clip,angle=0,height=\grauwidth]{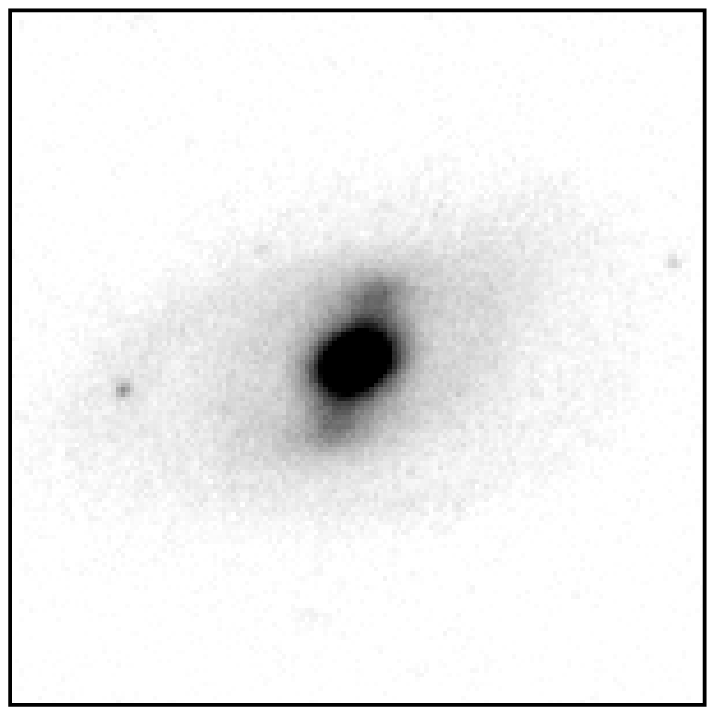}
\end{minipage}\\
\begin{minipage}[b][\greywidth][t]{4.7cm}
\includegraphics[bb = 95 613 314 777,clip,angle=0,height=\greywidth]{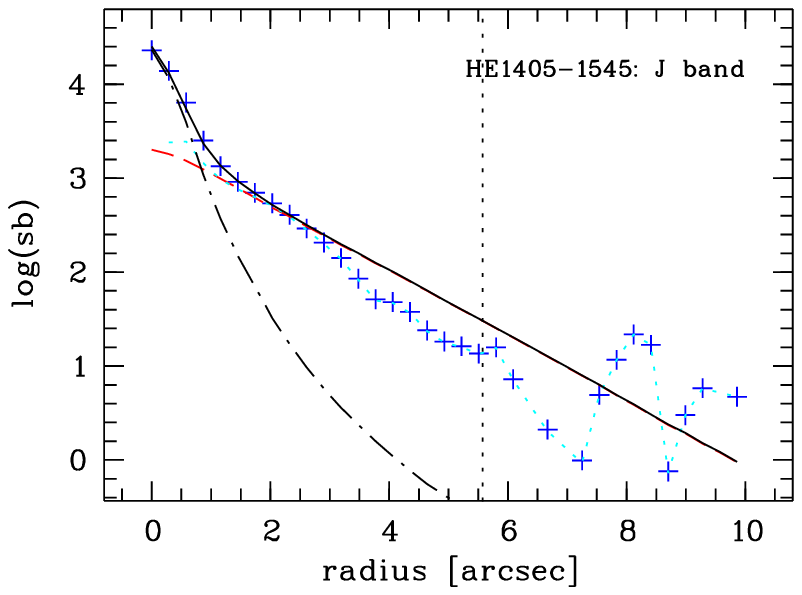}
\end{minipage}
\begin{minipage}[b][3.35cm][t]{\grauwidth}
\includegraphics[bb = 99 99 302 302,clip,angle=0,height=\grauwidth]{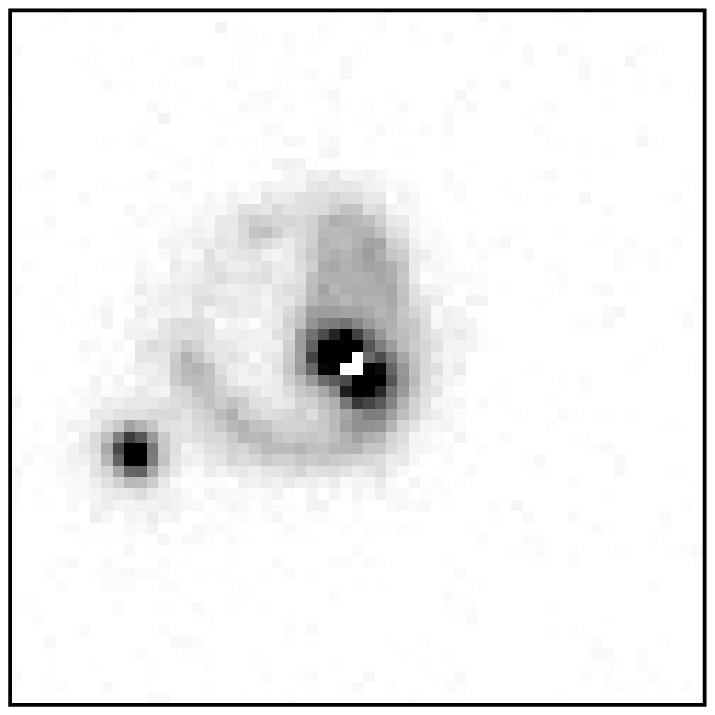}
\end{minipage}
\hfill
\begin{minipage}[b][\greywidth][t]{4.7cm}
\includegraphics[bb = 95 613 314 777,clip,angle=0,height=\greywidth]{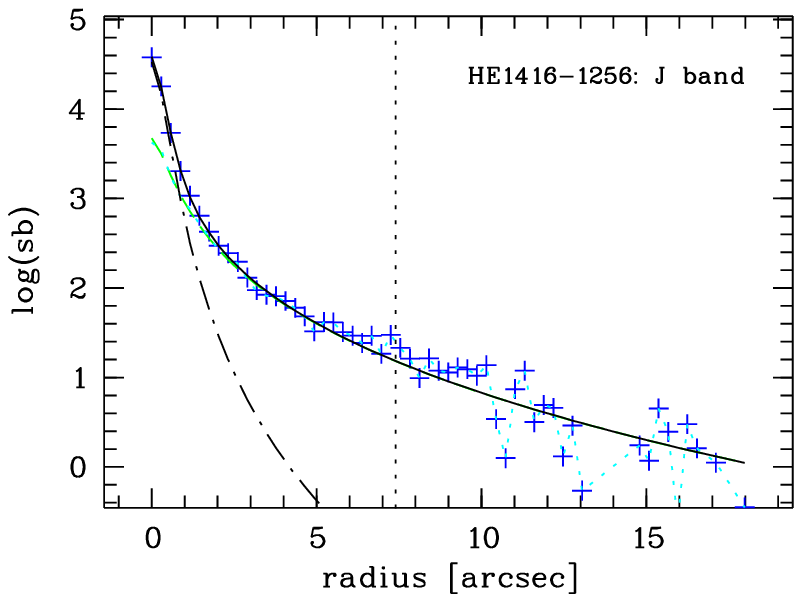}
\end{minipage}
\begin{minipage}[b][3.35cm][t]{\grauwidth}
\includegraphics[bb = 99 99 302 302,clip,angle=0,height=\grauwidth]{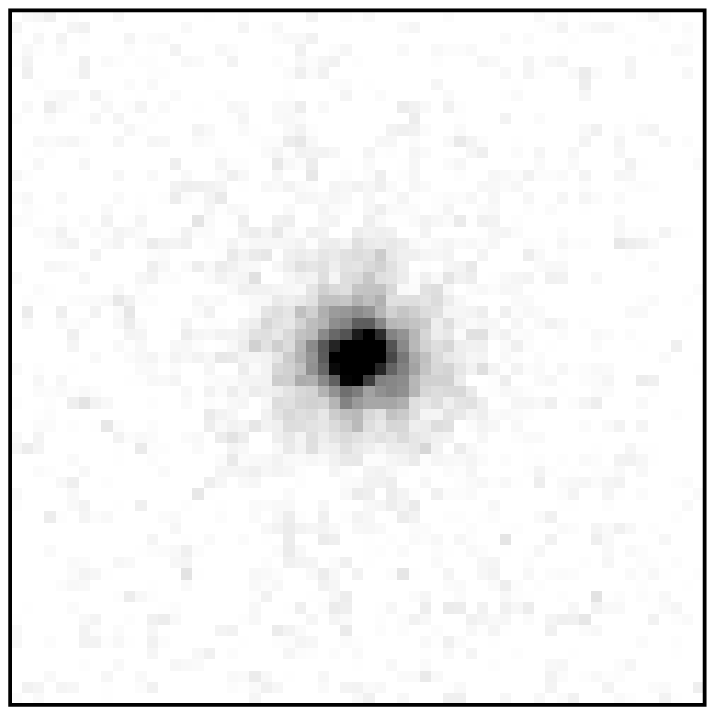}
\end{minipage}\\
\begin{minipage}[b][\greywidth][t]{4.7cm}
\includegraphics[bb = 95 613 314 777,clip,angle=0,height=\greywidth]{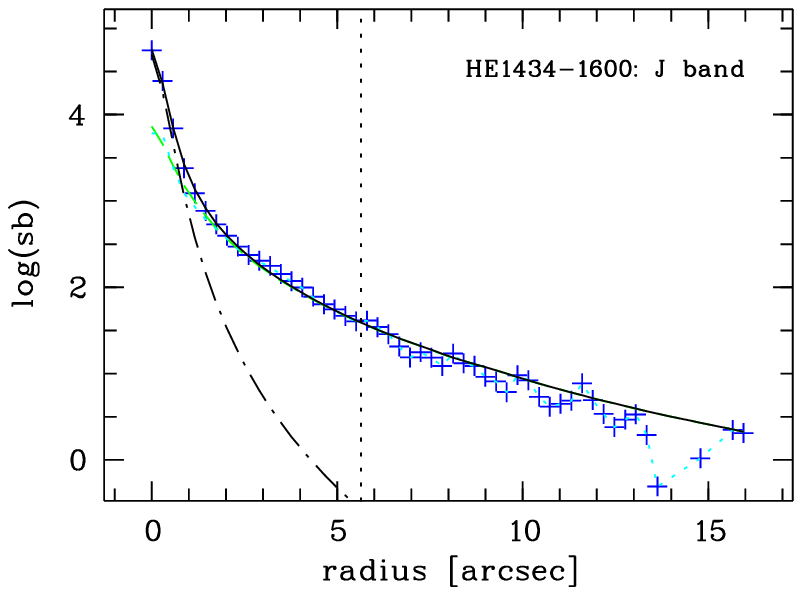}
\end{minipage}
\begin{minipage}[b][3.35cm][t]{\grauwidth}
\includegraphics[bb = 99 99 302 302,clip,angle=0,height=\grauwidth]{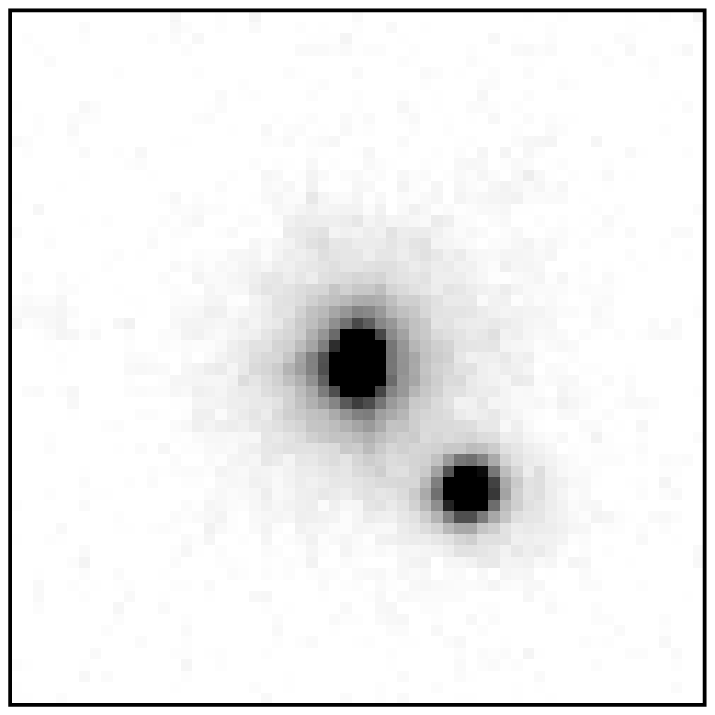}
\end{minipage}
\hfill
\begin{minipage}[b][1cm][t]{7cm}
\contcaption{}
\end{minipage}\\
\begin{minipage}[b][1cm][t]{\fullwidth}
\end{minipage}
\end{figure*}

\begin{table*}
\caption{\label{tab:objects2} 
Multi-component deblenging results. Shown are the general
morphological type of the host, (E)lliptical or (D)isk and in one case
a (B)ulge, half light radius $r_{1/2}$ in kpc, ellipticity $\epsilon$,
resulting inclination angles for the discs $i$, position angle from
north, and nucleus-to-host ratio N/H in restframe {\it V} and {\it
H}-band. Comments on morphology are given in the text for individual
objects.
}
\begin{center}
\begin{tabular}{lclllllll}
Object &
\multicolumn{1}{c}{Type}&
\multicolumn{1}{c}{$r_{1/2}$\,[kpc]}&
\multicolumn{1}{c}{$\epsilon$}&
\multicolumn{1}{c}{$i$\,[$^\circ$]}&
\multicolumn{1}{c}{$\varphi$\,[$^\circ$]}&
\multicolumn{1}{c}{N/H(V)}&
\multicolumn{1}{c}{N/H(H)}\\
\hline
HE\,0952--1552 &D & 6.8 &0.33 & 48.0&175.9&1.17 &0.59\\
HE\,1019--1414 &D & 5.0 &0.26 & 42.3&147.1&1.54 &1.36\\
HE\,1020--1022 &E &10.4 &0.28 &     &132.6&1.53 &1.28\\
HE\,1029--1401 &E & 4.2 &0.0  &     &--    &3.86 &2.53\\
HE\,1043--1346 &D & 6.7 &0.32 & 47.2&102.7&0.25 &0.42\\
HE\,1110--1910 &E & 5.6 &0.24 &     & 14.2&1.39 &1.33\\
HE\,1201--2409 &E & 1.5 &0.29 &     &  8.4&0.50 &0.66\\
HE\,1228--1637 &E & 3.5 &0.15 &     & 62.6&1.41 &1.57\\
HE\,1237--2252 &D & 9.9 &0.16 & 32.9& 39.4&0.34 &0.28\\
HE\,1239--2426 &D & 9.2 &0.22 & 38.7&162.7&0.48 &0.85\\
HE\,1254--0934 &D &16.0 &0.28 & 43.9& 95.5&6.18 &4.77\\
HE\,1300--1325 &E & 4.9 &0.39 &     & 34.0&0.36 &0.19\\
HE\,1310--1051 &D & 3.5 &0.11 & 27.1& 88.7&1.54 &1.21\\
HE\,1315--1028 &D & 8.1 &0.47 & 58.0&  1.3&1.03 &0.93\\
HE\,1335--0847 &E & 5.0 &0.19 &     & 22.0&0.57 &0.43\\
HE\,1338--1423 &D &13.7 &0.43 & 55.2&105.0&0.93 &0.31\\
& 		B & 4.6 &0.33 &     &145.3&0.93 &0.31\\
HE\,1405--1545 &D & 8.8 &0.35 & 49.5&165.0&1.28 &1.11\\
HE\,1416--1256 &E & 6.4 & 0.0 &     &   --&1.26 &1.12\\
HE\,1434--1600 &E & 7.2 &0.15 &     & 3.5 &1.44 &1.04\\
\hline
\end{tabular}
\end{center}
\end{table*}

In the {\it VRIJHK} bands the deblending was successful and
straightforward for all objects. The EFOSC $B$-band data were more
troublesome and could in general only be modelled with presetting the
scale lengths $r_\mathrm{1/2}$ from the other bands. Generally, the
integrated flux $F_\mathrm{host}$ of the host galaxy is a well
constrained parameter, in 2d multi-component deblending. However,
$r_\mathrm{1/2}$ alone is less well constrained and strongly
correlated with central surface brightness of the host
\citep{abra92,tayl96}. With $r_\mathrm{1/2}$ fixed, all but seven of
the 19 objects could be modelled also in $B$. In these seven cases the
host galaxy can be seen, but the residuals are too large to reliably
extract fluxes or morphological parameters. These frames were left out
from further analysis.

For 15 of our 133 frames, most of them in the $B$-band, the fixed
$r_\mathrm{1/2}$ fit ended with a host model systematically under- or
overestimating the data, which was clearly visible in the radial
profiles. In these cases the fit was disturbed by imperfections in the
PSF definition. A residual PSF mismatch was present in most of these
15 images. We corrected this by rescaling the resulting host models to
fit the QSO at radii where the influence of the nucleus was negligible
and reran the deblending with the host model fixed, to receive the best
fitting model of the nucleus for this constellation.

\subsection{Wavelength dependent scale lengths}\label{sec:mcvarscale}
For deblending the nucleus and host galaxy surface brightness
distributions we assumed that the scale lengths $r_{1/2}$ are constant
for all bands. Contrary to this assumption, studies of inactive
galaxies found significantly varying scale lengths for intermediate
and late type spirals \citep[e.g.][]{grij98, moel01}. The dominant
contribution comes from ongoing star formation, producing a population
of young, blue stars with a $M$/$L$ ratio much smaller than for the
dynamically dominant old population. The star formation occurs with a
more spread-out distribution, much different from the old
population. Thus compared to NIR bands tracing the distribution of old
stars, the $B$ and $V$-band will have an additional contribution of
the young population. In comparison a spiral will, at larger radii,
show more flux in the optical than in the NIR, and $r_{1/2}$ will
decrease with wavelength.

\Citet{grij98} studied a multicolour sample to describe morphological
properties of normal galaxies from ellipticals to late type
spirals. For early type galaxies up to Sa he found no change in
$r_{1/2}$ from $B$ to $K$. This changes for later types. For Sb the
ratio $r_{1/2}(B)/r_{1/2}(K)=1.3$, for Sc it is 1.6, but the spread is
always large.

We observe a significant trend only for HE\,1043--1346 and possibly
for HE\,1335--0847. In the latter case the deblending was inconclusive
to a degree that neither a pure exponential disc nor a spheroidal
model fitted all bands. While a spheroidal model agrees very well with
the profile in $V$, the fit is poor for radii $>6''$ in the
NIR. However, an exponential disc would fit quite well between $3''$
and $6''$, but it would underestimate the data at larger radii for
all bands except $K$.

For HE\,1043--1346 the case is much clearer, as we see resolved spiral
arms typical for a grand design Sc spiral (a grey scale plot is shown
in Fig.~\ref{fig:bb_grey1}). The colours of the arms are significantly
bluer than for the centre; while prominent in $B$ and $V$, they are
hardly detectable in $H$ and $K$. This produces exactly the effect
described above. When constraining the deblending region to $r<8''$
(27~kpc) -- as we did --, changes in the profile inside this radius
are visible, but small. The subtraction of the nuclear component is
biased only by a small amount. When on the other hand the spiral arms
were included, $r_{1/2}$ would change significantly.

The photometry as described in Section~\ref{sec:mcphot} is also
sensitive to varying scale lengths, because this corresponds to a
changing shape of the curve of growth of the host. The NIR bands with
smaller $r_{1/2}$ reach a given fraction of their total light at
smaller radii than in blue bands. To avoid a bias in photometry for
the two objects, we moved the radius $r_\mathrm{eq}$ to larger radii
than in the other cases. For HE\,1043--1346 in addition we extracted
the optical and NIR band separately to account for different shape. In
total we estimate the remaining systematic error between $B$ and $K$
band to be less than 5 per cent in flux.

Because of the sign of the potential change, $r_{1/2}$ {\it
decreasing} with wavelength, the fixed scalelength in the deblending
process would make the resulting colours {\it bluer}, whereas the bias
in photometry would make them systematically {\it redder}. Since the
effect of deblending is negligible except for the two objects
mentioned, the colours discussed in section~\ref{sec:mccolours} are
not biased towards bluer values by these effects.

\subsection{Morphology}\label{sec:mcresultmorph}
\begin{figure}
\includegraphics[bb = 50 53 313 314,clip,angle=0,width=\colwidth]{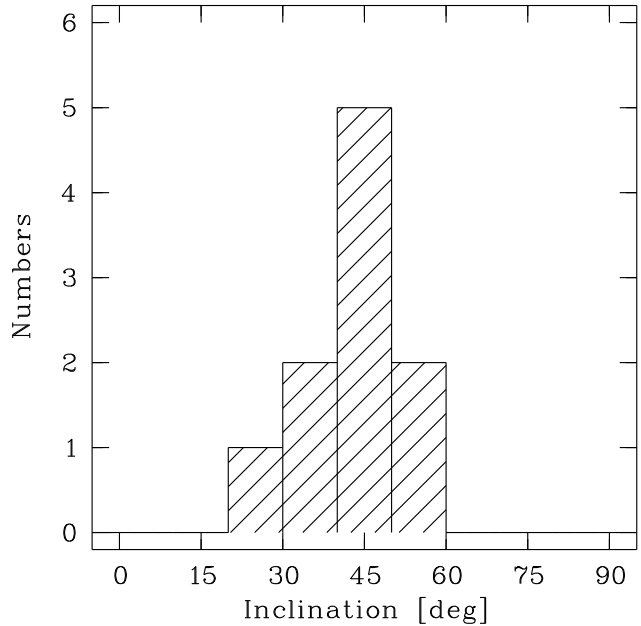}
\caption{\label{fig:inclinations}Inclinations of the ten discs in the
sample. Apparent is the lack of high inclinations $i> 60^\circ$
conforming to the unified AGN scheme.
}
\end{figure}

\begin{figure}
\includegraphics[bb = 50 53 313 314,clip,angle=0,width=\colwidth]{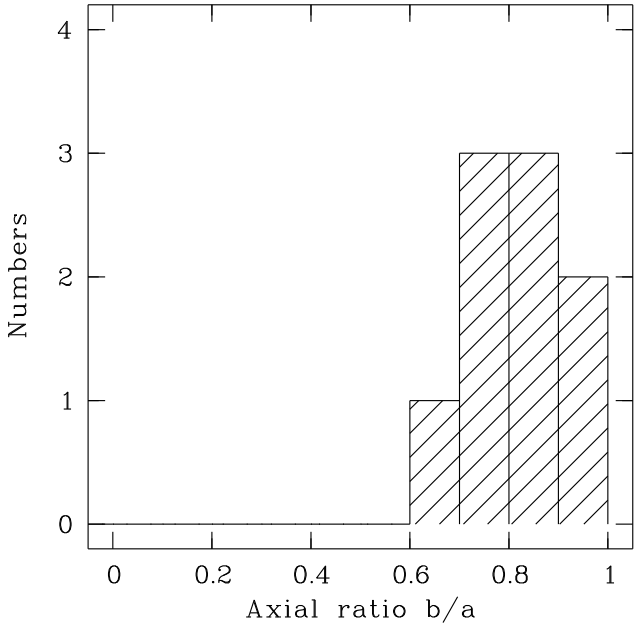}
\caption{\label{fig:ellipticities}Axial ratios of the nine ellipticals
in the sample. No host exists more elliptic than E4. Average and
distribution are very similar to the general population of bright
inactive ellipticals.}
\end{figure}

Of the 19 objects, nine were classified from the best fitting models
as being disc dominated, nine as spheroid dominated and one object as
a disc plus bulge. A summary of the morphological parameters is given
in Table~\ref{tab:objects2}. Scale lengths range from 3.5 to 16.0\,kpc
for the discs, with an average of 8.8\,kpc, and from 1.5 to 10.4\,kpc
for the elliptical hosts, with an average of 5.4\,kpc. Mean
ellipticities are 0.19 for the ellipticals and 0.29 for the
discs. Since we mostly used only a single component for the host
galaxy, we lack information about the relation between bulge and disc
luminosities. Thus we do not know about the disc thicknesses and
cannot give inclinations as originally defined by \citet{hubb26}. We
estimate inclinations $i$ assuming the discs to be thin,
$i=\arccos(b/a)=\arccos(1-e)$. With this definition the average
inclination for the discs in the sample is 44.8$^\circ$. The
distribution is plotted in Fig.~\ref{fig:inclinations}. For the
ellipticals we give the distribution of axial ratios $q=b/a=1-e$ in
Fig.~\ref{fig:ellipticities}.

The residual PSF mismatch and the presence of individual features
(knots, weak unmasked foreground stars, etc.) have an influence on
ellipticities, potentially biasing the deblending routine against
values close to zero. A feature like a PSF mismatch or a foreground
object will generally show a non-circular symmetry. When fitted -- by
force -- with elliptical isophotes, the feature will therefore display
a non-zero ellipticity. When the $\chi^2$ minimisation attempts to
model the feature geometry by adjusting host model parameters
accordingly, the ellipticities of the host are changed. The resulting
host does not need to show any apparent fault, nor does a slight
mismatch have a significant influence on the flux of the nuclear
model, but ellipticities will be biased away from zero.

We find this to be the case for two objects. HE\,1029--1401 and
HE\,1416--1256 were both assigned ellipticities between 0.15 and 0.2
in the free parameter fit, even though their appearance on the
detector is very circular. For the fixed parameter fit we therefore
forced $e=0$. Both objects are clear elliptical galaxies and we do not
find similar cases for the discs in our sample. Thus we conclude that
the absence of face-on discs with $i< 25^\circ$ ($e<0.1$) in the
sample is real (Fig.~\ref{fig:inclinations}). Also missing are high
inclination systems with $i> 60^\circ$ ($e>0.5$) as expected from AGN
unification models.

\medskip

\noindent We now give some comments on the morphological
characteristics of the 19 objects. Spectral information about
association of companions is available from our own unpublished
spectroscopy:
\medskip

\noindent{\bfseries HE\,0952--1552}: Isolated disc galaxy with
$r_\mathrm{1/2} = 6.8 \mkpc$ and $i=48^\circ$. An isophote twist in
the NIR suggests spiral arms, but no arms are resolved. Otherwise very
symmetric.
\medskip

\noindent{\bfseries HE\,1019--1414}: Disc with $r_\mathrm{1/2} = 5.0
\mkpc$ and $i=42^\circ$. No visible spiral arms. An asymmetry to the
north is leading into a prominent tidal arm that was excluded from
photometry, extending $10''$ or 20~kpc to the north. A knot in this
arm is a companion galaxy about 2~mag fainter in $V$ than the host.
\medskip

\noindent{\bfseries HE\,1020--1022}: This is the most distant object
in the sample, at $z=0.197$ and with $r_\mathrm{1/2} = 10.4 \mkpc$
also the largest elliptical, and in the NIR also the most luminous. It
appears very symmetric, with no signs of distortions. A luminous
galaxy exists 90~kpc projected distance away to the north-west.
\medskip

\noindent{\bfseries HE\,1029--1401}: One of the brightest QSOs in the
sky, and showing the most luminous nucleus in the sample. The host is
nearly circular, very luminous elliptical that can be traced out to
$>35\mkpc$ from the nucleus, with $r_\mathrm{1/2} = 4.2 \mkpc$. A weak
`blob' 16~kpc to the north and a galaxy 50~kpc to the north are
confirmed physical companions. This object suffers from a very poor
quality of the available PSF stars and a very bright nucleus, both
increasing the uncertainty of the deblending.
\medskip

\noindent{\bfseries HE\,1043--1346}: This objects shows rather open
spiral arms, typical of an Sb or Sc spiral, plus a bar-like structure
surrounding the nucleus. For nucleus removal only the inner 12~kpc
were fitted, corresponding to the bar. It follows clearly an
exponential disc law with $r_\mathrm{1/2} = 6.7 \mkpc$ and ellipticity
$e=0.32$. The spiral arms further out are less elongated,
corresponding to an inclination angle of $i=34^\circ$ or
$e=0.17$. This spiral is isolated, with no nearby galaxies or other
companions visible.
\medskip

\noindent{\bfseries HE\,1110--1910}: This host is a rather symmetric
elliptical galaxy, with $r_\mathrm{1/2} = 5.6 \mkpc$ and $e=0.24$. A
galaxy to the north-west is a confirmed companion. It is bluer than
the host in $V-K$ by about 0.5~mag.
\medskip

\noindent{\bfseries HE\,1201--2409}: Symmetric and most compact
elliptical in the sample with $r_\mathrm{1/2} = 1.5 \mkpc$, in $B$ the
host is barely resolved. No further structure or interaction is
visible. Around the object a group of galaxies is visible, at 80, 86,
134, 166 and 205~kpc projected distance. No redshifts exist for these
galaxies, thus the physical association is not confirmed.
\medskip

\noindent{\bfseries HE\,1228--1637}: Elliptical host with
$r_\mathrm{1/2} = 3.5 \mkpc$ and a slight east--west asymmetry. It lies in
the vicinity of a group of galaxies with 37, 75, 175 and 235~kpc
projected distance, but no tidal connection is visible. The closest
galaxy to the east is a confirmed companion at the same redshift. No
information is available on the association of the remaining galaxies.
\medskip

\noindent{\bfseries HE\,1237--2252}: Large disc with $r_\mathrm{1/2} =
9.9 \mkpc$ and $i=33^\circ$. An armlike asymmetry to south is visible
in all bands and appears to be of tidal origin.
\medskip

\noindent{\bfseries HE\,1239--2426}: This is a more complex object
with a central bulge or a bar and very tightly wound arms, almost a
ring. At least one bright knot in the ring is visible from $B$ to
$K$. A two-component host fit was not successful, thus we stayed with
a one component disc model that fits the inner parts better than a
spheroidal model. The parameters thus only describe the inner part
with $r_\mathrm{1/2} = 9.2 \mkpc$ and $e=0.22$. The ringlike outer
structure has the same orientation as the inner part, and slightly
lower ellipticity of $e=0.17$, corresponding to $i=34^\circ$. A
confirmed associated companion is located at only 26~kpc to the
north-west.
\medskip

\noindent{\bfseries HE\,1254--0934}: Complex interacting system of the
QSO and at least one, possibly two other galaxies. The host shows an
exponential disc profile, with $r_\mathrm{1/2} = 16.0 \mkpc$ and
ellipticity $e=28$. The first companion is of similar luminosity as
the host, 14~kpc away to the west. A tidal arm extends in an arc to
the south and west towards the second companion at 60~kpc distance,
1.3~mag fainter than the host in $V$. About 460~kpc projected distance
to the east exists a small group of five bright galaxies. Both
companions and the tidal arm were masked in the deblending.
\medskip

\noindent{\bfseries HE\,1300--1325}: Interacting galaxy classified as
an elliptical with $r_\mathrm{1/2} = 4.9 \mkpc$ and $e=0.39$. The
morphology is complex and asymmetric to the centre, thus $e$ might be
slightly overestimated and more likely have a value around 0.3. The
host is interacting with a luminous galaxy 43~kpc to the east. Both
galaxies are tidally disturbed, the host is displaying two tidal arms
and a tidal bridge exists between the two galaxies.
\medskip

\noindent{\bfseries HE\,1310--1051}: Clear disc with two asymmetric
spiral or tidal arms, $r_\mathrm{1/2} = 3.5 \mkpc$,
$i=27^\circ$. There is no luminous companion visible as a source for
tidal interaction, the closest extended objects are a galaxy 0.5~mag
fainter than the host in $V$, 45~kpc to the north, and a second,
110~kpc east, 2~mag fainter, but their redshifts are not known.
\medskip

\noindent{\bfseries HE\,1315--1028}: Disc with $r_\mathrm{1/2} = 8.1
\mkpc$, $i=58^\circ$, but the latter seems slightly overestimated
judging from the produced model. The host is asymmetric in the
north-west. An also asymmetric companion exists 82~kpc in the same
direction, 0.6~mag fainter than the host in $V$.
\medskip

\noindent{\bfseries HE\,1335--0847}: A decision between disc and
spheroidal was difficult in this case. In the range between $3''$ and
$6''$ an exponential disc would also be a good fit, so we could have a
disc with additional flux at larger radii from star formation. This
excess decreases somewhat from $V$ to the NIR. The results were
inconclusive and in total we opted for the spheroid model. The host is
very symmetric, with $r_\mathrm{1/2} = 5.0 \mkpc$, $e=0.19$. No
companions or galaxies in the vicinity are visible.
\medskip

\noindent{\bfseries HE\,1338--1423}: Prominent spiral with bulge or
bar. The disc was modelled with $r_\mathrm{1/2} = 13.7 \mkpc$,
$i=55^\circ$, the bulge/bar with a spheroid with $r_\mathrm{1/2} = 4.6
\mkpc$, $e=0.33$, but other geometries for the bulge/bar are possible.
\medskip

\noindent{\bfseries HE\,1405--1545}: Highly disturbed disc with tidal
arms to the north and south. There is a luminous knot 24~kpc
south-east of the nucleus, touched by the southern arm, that could be
an interacting companion. The morphological parameters are
$r_\mathrm{1/2} = 8.8 \mkpc$, $i=49.5^\circ$. The deblending of this
object was very difficult. It excluded the complete north-eastern half
of the frame and left significant residuals.
\medskip

\noindent{\bfseries HE\,1416--1256}: Very symmetric circular
elliptical with $r_\mathrm{1/2} = 6.4 \mkpc$. The closest galaxies are
at 150, 180, and 240~kpc distance.
\medskip

\noindent{\bfseries HE\,1434--1600}: Symmetric elliptical with low
ellipticity. A companion 2~mag fainter in $V$ is located 14~kpc to the
south-west.
\medskip

\subsection{Colours}\label{sec:mccolours}

\begin{table*}
\caption{\label{objects3}
Apparent and absolute magnitudes for the host galaxies. Conversion to
absolute magnitudes uses the {\it K}-corrections from
Table\,\ref{t_kcorr}. In $B$ no value is given if the deblending
process failed to deliver satisfactory results.
}
\begin{center}
\begin{tabular}{lcccccccccccccc}
Object
&\multicolumn{1}{c}{$B$}
&\multicolumn{1}{c}{$V$}
&\multicolumn{1}{c}{$R$}
&\multicolumn{1}{c}{$I$}
&\multicolumn{1}{c}{$J$}
&\multicolumn{1}{c}{$H$}
&\multicolumn{1}{c}{$K$}
&\multicolumn{1}{c}{$M_B$}
&\multicolumn{1}{c}{$M_V$}
&\multicolumn{1}{c}{$M_R$} 
&\multicolumn{1}{c}{$M_I$}
&\multicolumn{1}{c}{$M_J$} 
&\multicolumn{1}{c}{$M_H$}
&\multicolumn{1}{c}{$M_K$}\\
\hline										     
HE\,0952--1552 &17.6 &16.7 &16.1 &15.6 &14.5 &13.9 &13.4 & --21.9& --22.5& --23.1& --23.6& --24.6& --25.2& --25.5\\
HE\,1019--1414 &18.0 &17.2 &16.5 &16.1 &15.2 &14.5 &14.1 & --20.6& --21.3& --21.9& --22.4& --23.2& --23.9& --24.1\\
HE\,1020--1022 &19.2 &18.0 &17.2 &16.5 &15.6 &14.9 &14.3 & --22.2& --22.9& --23.5& --24.1& --24.9& --25.5& --25.8\\
HE\,1029--1401 &   * &15.6 &15.2 &14.7 &14.1 &13.5 &12.3 &      *& --23.2& --23.4& --23.9& --24.5& --25.1& --25.8\\
HE\,1043--1346 &16.6 &15.9 &15.4 &14.9 &14.0 &13.3 &13.0 & --21.7& --22.3& --22.8& --23.3& --24.1& --24.8& --24.9\\
HE\,1110--1910 &17.9 &17.2 &16.5 &16.0 &15.4 &14.7 &14.1 & --21.8& --22.2& --22.8& --23.2& --23.8& --24.5& --24.8\\
HE\,1201--2409 &17.9 &16.9 &16.3 &16.0 &15.5 &14.8 &14.0 & --22.5& --23.0& --23.5& --23.8& --24.2& --24.8& --25.3\\
HE\,1228--1637 &17.7 &16.9 &16.3 &16.0 &15.1 &14.5 &14.1 & --21.8& --22.3& --22.8& --23.1& --24.0& --24.5& --24.6\\
HE\,1237--2252 &17.1 &16.3 &15.7 &15.2 &14.3 &13.7 &13.3 & --22.1& --22.7& --23.2& --23.7& --24.6& --25.2& --25.4\\
HE\,1239--2426 &16.7 &16.1 &15.5 &15.1 &14.0 &13.5 &13.1 & --22.1& --22.5& --23.0& --23.5& --24.5& --25.0& --25.2\\
HE\,1254--0934 &   * &17.3 &16.9 &16.0 &15.3 &14.8 &13.9 &      *& --22.5& --22.9& --23.7& --24.4& --24.9& --25.5\\
HE\,1300--1325 &   * &15.3 &14.7 &14.2 &13.3 &12.7 &12.4 &      *& --22.1& --22.6& --23.1& --24.0& --24.6& --24.8\\
HE\,1310--1051 &16.5 &16.0 &15.5 &15.0 &14.1 &13.6 &13.3 & --20.2& --20.6& --21.1& --21.6& --22.4& --23.0& --23.2\\
HE\,1315--1028 &   * &17.7 &17.0 &16.4 &15.4 &14.9 &14.5 &      *& --21.3& --22.0& --22.6& --23.5& --24.1& --24.2\\
HE\,1335--0847 &17.5 &16.8 &16.3 &15.9 &15.1 &14.5 &14.3 & --21.3& --21.8& --22.2& --22.6& --23.4& --23.9& --23.9\\
HE\,1338--1423 &15.1 &14.5 &14.0 &13.4 &12.5 &11.8 &11.4 & --22.0& --22.6& --23.0& --23.6& --24.5& --25.2& --25.4\\
HE\,1405--1545 &   * &17.4 &16.9 &16.3 &15.6 &15.0 &14.4 &      *& --23.3& --23.7& --24.2& --24.9& --25.4& --25.6\\
HE\,1416--1256 &   * &17.5 &17.0 &16.3 &15.9 &15.2 &14.7 &      *& --22.3& --22.6& --23.3& --23.7& --24.3& --24.5\\
HE\,1434--1600 &   * &17.0 &16.4 &15.8 &15.1 &14.4 &13.7 &      *& --23.1& --23.5& --24.1& --24.8& --25.3& --25.7\\
\end{tabular}
\end{center}
\end{table*}

\begin{table*}
\caption{\label{objects4}
Host galaxy rest-frame colours after application of the {\it
K}-corrections from Table\,\ref{t_kcorr}.  For $B-V$ no value is given
if the deblending process failed to deliver satisfactory results.
}
\begin{center}
\begin{tabular}{lccccccc}
Object  
&\multicolumn{1}{c}{$B-V$}
&\multicolumn{1}{c}{$V-R$} 
&\multicolumn{1}{c}{$R-I$}
&\multicolumn{1}{c}{$I-J$} 
&\multicolumn{1}{c}{$J-H$}
&\multicolumn{1}{c}{$H-K$} 
&\multicolumn{1}{c}{$V-K$}\\
\hline
HE\,0952--1552 &0.57 &0.64 &0.47 &1.04 &0.60 &0.22 &2.96\\
HE\,1019--1414 &0.62 &0.68 &0.42 &0.86 &0.64 &0.24 &2.84\\
HE\,1020--1022 &0.68 &0.58 &0.65 &0.78 &0.61 &0.22 &2.84\\
HE\,1029--1401 &   * &0.27 &0.48 &0.63 &0.55 &0.71 &2.64\\
HE\,1043--1346 &0.57 &0.54 &0.48 &0.86 &0.63 &0.19 &2.70\\
HE\,1110--1910 &0.41 &0.62 &0.39 &0.59 &0.66 &0.34 &2.61\\
HE\,1201--2409 &0.57 &0.47 &0.26 &0.44 &0.65 &0.48 &2.29\\
HE\,1228--1637 &0.48 &0.49 &0.35 &0.84 &0.49 &0.19 &2.35\\
HE\,1237--2252 &0.61 &0.51 &0.47 &0.95 &0.56 &0.17 &2.66\\
HE\,1239--2426 &0.48 &0.50 &0.47 &0.98 &0.53 &0.21 &2.70\\
HE\,1254--0934 &   * &0.38 &0.78 &0.68 &0.49 &0.60 &2.92\\
HE\,1300--1325 &   * &0.47 &0.49 &0.92 &0.63 &0.20 &2.70\\
HE\,1310--1051 &0.47 &0.48 &0.46 &0.86 &0.53 &0.20 &2.53\\
HE\,1315--1028 &   * &0.67 &0.54 &0.96 &0.53 &0.12 &2.83\\
HE\,1335--0847 &0.47 &0.42 &0.43 &0.74 &0.52 &0.05 &2.14\\
HE\,1338--1423 &0.56 &0.43 &0.65 &0.88 &0.65 &0.28 &2.89\\
HE\,1405--1545 &   * &0.43 &0.52 &0.63 &0.55 &0.17 &2.31\\
HE\,1416--1256 &   * &0.33 &0.64 &0.41 &0.58 &0.19 &2.16\\
HE\,1434--1600 &   * &0.34 &0.64 &0.65 &0.58 &0.38 &2.59\\
\end{tabular}
\end{center}
\end{table*}

The apparent and absolute magnitudes of the host galaxies of our
objects in the seven bands are collected in Table~\ref{objects3}. From
these we derived the rest-frame colours in Table~\ref{objects4} which
include the $K$-correction terms from Table~\ref{t_kcorr}. Since the
$B$-band images have on average much lower S/N than the data from the
other bands, and also data is not available for all objects, we
include $V-K$ as a long wavelength baseline colour for the discussion
below. In Table~\ref{cols_subsamples}, mean and median colours are
listed for the sample as a whole and for subsamples of ellipticals and
discs.

The main internal error source for the colours are the uncertainties
of the deblending process. Simulations as performed by \citet{kuhl02}
suggest that deblending errors for our objects are of the order of
0.05~mag in host galaxy magnitude, and rather 0.1~mag for the very
compact elliptical HE\,1201--2409.

\begin{table*}
\caption{\label{cols_subsamples}
Colours of the sample as a whole and for subsamples of hosts
identified as discs or as ellipticals. Listed are mean with error of
mean, and median values in {\it italics}.
}
\begin{center}
\begin{tabular}{lccccccc}
 &\multicolumn{1}{c}{$B-V$}&\multicolumn{1}{c}{$V-R$}&\multicolumn{1}{c}{$R-I$}&\multicolumn{1}{c}{$I-J$}&\multicolumn{1}{c}{$J-H$}&\multicolumn{1}{c}{$H-K$}&\multicolumn{1}{c}{$V-K$}\\
\hline

Whole sample	&0.54	&0.49	&0.50	&0.77	&0.58	&0.27	&2.61\\
		&(0.02)	&(0.03)	&(0.03)	&(0.04)	&(0.01)	&(0.04)	&(0.06)\\
		&{\it 0.56}&{\it 0.47}&{\it 0.47}&{\it 0.78}&{\it 0.56}&{\it 0.20}&{\it 2.64}\\
Ellipticals	&0.52	&0.44	&0.48	&0.67	&0.59	&0.31	&2.48\\
		&(0.05)	&(0.04)	&(0.05)	&(0.06)	&(0.02)	&(0.07)	&(0.08)\\
		&{\it 0.47}&{\it 0.42}&{\it 0.43}&{\it 0.63}&{\it 0.58}&{\it 0.20}&{\it 2.35}\\
Discs		&0.55	&0.53	&0.53	&0.87	&0.57	&0.24	&2.73\\
		&(0.02)	&(0.03)	&(0.04)	&(0.04)	&(0.02)	&(0.04)	&(0.06)\\
		&{\it 0.56}&{\it 0.50}&{\it 0.47}&{\it 0.86}&{\it 0.55}&{\it 0.20}&{\it 2.70}\\
\hline
\end{tabular}
\end{center}
\end{table*}

\begin{figure}
\includegraphics[bb = 40 52 313 314,clip,angle=0,width=\colwidth]{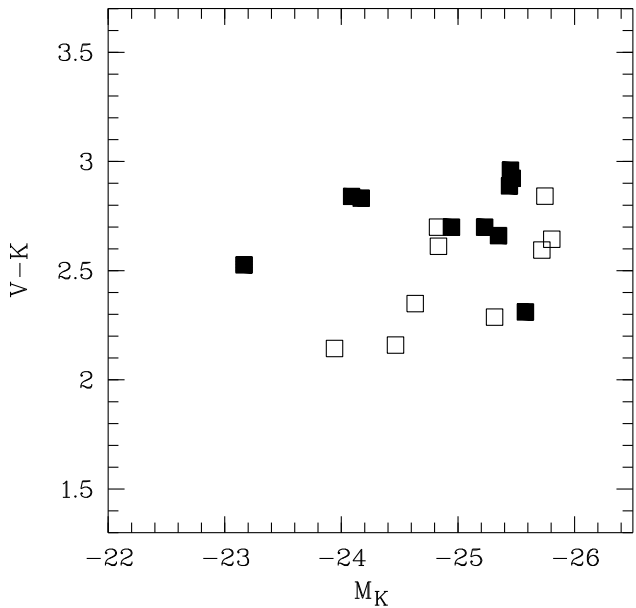}
\caption{\label{fig:vh_disk_ellipse} $V-K$ colours against $K$-band
absolute magnitude $M_K$ of the individual hosts in the multicolour
sample. Open symbols mark ellipticals, filled symbols discs.  }
\end{figure}

\subsection{Influence of emission lines}\label{emissionlines}
Emission lines from the ISM of the host galaxy can influence the broad
band colours. In two samples of host galaxy spectra
\citep{cour02b,jahn02}, we find a significant fraction of host
galaxies showing strong extended gas emission lines, particularly
prominent for elliptical host galaxies.

For $z<0.2$ the $V$ broad band filter contains H$\beta$ and the
[O\,III] doublet around 5000~\AA. We estimate the line fluxes to
correspond to an equivalent of 0.0--0.15~mag of the $V$-band continuum
emission. Similarly, for H$\alpha$ and the [N\,II] doublet the $I$-band
might show enhanced flux for objects with $0.1<z<0.35$, and to a
smaller extend also the $R$-band for $z<0.1$.

Furthermore, we find in the mentioned studies that the gas in the
spheroids is mainly present in rotating gas discs, implying that the
emission line regions have a different spatial distribution than the
stars. For the deblending this can result in an additional
overestimation of the host galaxy flux, as a result of the algorithm
attempting to compensate for the slight model mismatch.

We estimate that the combined average bias from emission lines and
deblending errors for the $V$-band is of the order of $-$0.15~mag at
most. Judging from the radial profiles of our objects in the different
bands, we can exclude systematically larger deviations. A correction
for this bias has not been included in Tables~\ref{objects3} to
\ref{cols_sample}.

\section{Colour comparison to inactive galaxies}\label{sec:comp_inactive}

In Table~\ref{cols_subsamples} no significant difference can be seen
for the colours of neighbouring wavelength bands of the discs and
ellipticals in the sample. For the long baseline $V-K$ however a
0.25~mag (median: 0.35~mag) difference exists, the ellipticals being
{\it bluer} than the discs. This is not created by statistical
outliers, as Figure~\ref{fig:vh_disk_ellipse} illustrates: The
ellipticals (open symbols) have systematically bluer colours than the
discs (filled symbols). Some of the difference might be caused by line
emission as discussed above. If for this a 0.15~mag correction is
applied to the $V-K$ colours of the ellipticals, their average colour
increases to $V-K=2.63$ and the difference to the discs in the sample
decreased to 0.1~mag. This is not a statistically significant
difference, as confirmed by Student's $t$ test. Thus the ellipticals and
discs in the sample have similar $V-K$ colours. While the comparison
of ellipticals and discs in the sample is already illustrative, we are
rather interested to relate our results to the general population of
{\it inactive} galaxies.

\subsection{Empirical colours of inactive galaxies}\label{sec:emp_inaccols}
Optical colours of normal galaxies are well established. Based on
systematic spectrophotometric catalogues
\citep[e.g.][]{kenn92,kinn96}, broad band colours of galaxies can be
computed, and average values predicted. In the UV and optical,
\citet{fuku95} list average colours of inactive galaxies for 48
photometric bands. For ellipticals their numbers agree down to about
0.05\,mag with the effective colours derived by \citet{mann01}, who
compiled colours for galaxies with $M_V<-21$ from a large number of
publications (but not from Fukugita et al.). Since {\it effective}
colours are computed from fluxes inside an aperture with constant {\it
radius} for all bands, a comparison with total colours is only
possible for ellipticals, showing small radial colour gradients.

In a statistically extensive study of more than 1000 galaxies from
catalogue data, \citet{fioc99} computed total and effective NIR and
optical--NIR colours, using type-dependent curves of growth for
converting aperture photometry data. The authors also investigate the
effect of disc inclination on $B-H$ and the colour-magnitude relation
differentiated by Hubble type. The values for NIR colours are almost
independent of type. They find $J-H=0.71$ and $H-K=0.20$ for
ellipticals, rising to 0.78 and 0.25 for Sb spirals, which is again
consistent with the values by Mannucci et al. Fioc \& Rocca-Volmerange
find an effective $B-H=3.86$ for ellipticals, which is about 0.2~mag
bluer than computed by \citet{mann01}.

We cannot resolve these discrepancies at this point. We will take the
colours listed by \citeauthor{fioc99} as a basis because a) the
definition of their total colours is closer to the methods used in our
study, b) because their sample is larger and the treatment seems very
consistent, and c) because their colours are slightly bluer, which is
the same direction of bias as all of our other potential systematic
errors. We thus get conservative estimates for the blue colours of our
sample.

We used their $B-H$ colour relation, including inclination and
magnitude dependencies, to calibrate the comparison optical--NIR
colours for inactive galaxies. Our disc subsample has on average an
axial ratio $\overline{R}=\overline{a/b}=1.4$, a value for which the
\citeauthor{fioc99} relations yield $B-H=3.34$. Including the
colour-magnitude dependence changes this value only slightly to
$B-H=3.30$, because the average brightness of their Sb and our disc
subsample are very close to each other (0.25~mag difference).  For
elliptical galaxies the total colours at the average brightness of our
subsample are $B-H=3.75$, corrected for magnitude dependence.

In Table~\ref{cols_sample} we list the collected colours as taken from
\citet{fuku95} for the optical colours, the optical--NIR and NIR
colours are computed from \citeauthor{fioc99}. For the discs we assume
intermediate type (Sb) colours.

\begin{table*}
\caption{\label{cols_sample} Disc and ellipticals in the sample and
comparison with inactive galaxies. For the discs intermediate type Sb
is used. Sources: optical colours from \citet{fuku95}, optical--NIR
and NIR colours computed from \citet{fioc99} as described in the
text. $I-J$ was calculated from the other colours. For $V-K$ the
spread is given in parentheses.  }
\begin{center}
\begin{tabular}{lccccccc}
 &\multicolumn{1}{c}{$B-V$}&\multicolumn{1}{c}{$V-R$}&\multicolumn{1}{c}{$R-I$}&\multicolumn{1}{c}{$I-J$}&\multicolumn{1}{c}{$J-H$}&\multicolumn{1}{c}{$H-K$}&\multicolumn{1}{c}{$V-K$}\\
\hline
Ellipticals\\
Inactive	&0.96	&0.61	&0.70	&0.77	&0.71	&0.20	&2.99 (0.12)\\
QSO host sample	&0.52	&0.44	&0.48	&0.67	&0.59	&0.31	&2.48 (0.25)\\
$\Delta$	&0.44	&0.17	&0.22	&0.10	&0.12	&--0.11	&0.50\\
\hline		 	 	 	 	 	 
Discs (Sb)\\
Inactive	&0.68	&0.54	&0.63	&0.67	&0.78	&0.25	&2.87 (0.36)\\
QSO host sample	&0.55	&0.53	&0.53	&0.87	&0.57	&0.24	&2.73 (0.20)\\
$\Delta$	&0.13	&0.01	&0.10	&--0.20	&0.21	&0.01	&0.14\\
\hline
\end{tabular}
\end{center}
\end{table*}

\subsection{Abnormally blue QSO host galaxies}\label{mc_colourcomparison}
In Table~\ref{cols_sample} we compare colours for inactive galaxies
with the mean colours of our subsamples (from
Table~\ref{cols_subsamples}). We find very blue colours for our
elliptical host galaxies in comparison, and slightly bluer colours
than normal for the discs. The long wavelength baseline colour $V-K$
is for our ellipticals 0.5~mag bluer (0.35~mag including correction
for $H\beta$ emission in $V$) than for inactive galaxies of similar
luminosity. The difference in $B-V$ is $\sim0.4$~mag. We again
performed a $t$ test with the hypothesis of identical mean $V-K$
colours for our ellipticals and their inactive counterparts. This
hypothesis is rejected at $>$99.9 per cent confidence level. We also tested
whether their colours are consistent with those of late type (Sc or
later) inactive galaxies showing strong star formation. This
hypothesis is not rejected by the $t$ test. The host galaxies
classified as discs are also, on average, slightly bluer than their
inactive counterparts, but not by a significant amount.

Are these colours real? The dependency of the scale length on
wavelength does not apply to the ellipticals, so all known systematic
effects in our procedure would only produce {\it redder} colours.

There is one additional possible artefact that could be created in the
deblending procedure. This is transfer of a \emph{constant} fraction
of nuclear light into the host, e.g., due to a similar systematic
error of host galaxy parameter estimation in all bands. This would
imply a contamination of the host galaxy with light from the
nucleus. Host galaxies with spheroidal morphology are in general more
compact than exponential discs, owing to their $r^{1/4}$ surface
brightness distribution. They are closer in their appearance to the
PSF shape of the nucleus than discs and thus modelling of the nuclear
contribution is generally more sensitive to the properties of the
host.

But, while for optical colours the nucleus is much bluer than any
host, for optical--NIR colours such a transfer would have the opposite
effect, owing to the fact that quasar nuclei are generally quite red
in $V-K$, with $(V-K)_\mathrm{nuc}=3.4$ \citep[cf.][]{elvi94}. This is
similar or even {\it redder} than for inactive galaxies and
particularly redder than our host galaxy sample! Thus a constant
fraction of nuclear flux transfered to the host would not make $V-K$
bluer, but redder.

\begin{figure}
\includegraphics[bb = 40 53 313 314,clip,angle=0,width=\colwidth]{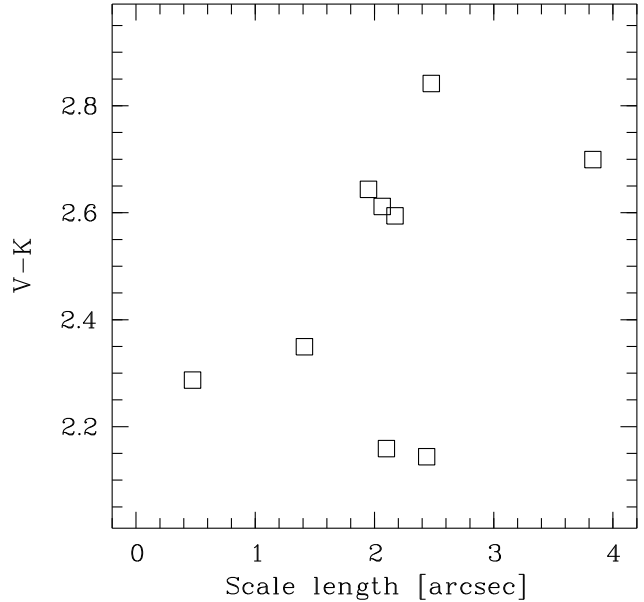}
\caption{\label{fig:mc_vk_compactness} 
$V-K$ colours for the ellipticals in the sample vs.\ scale length in
arcsec as a measure of compactness in the image. No trend, as an
indication for systematic flux transfer depending on compactness, can
be seen.
}
\end{figure}

In fact, we can even exclude significant flux transfer for our
deblended data even for the more compact objects. There is no
correlation between compactness of the host (as parametrised by the
angular scale length) and the $V-K$ colour of the host galaxy
(Fig.~\ref{fig:mc_vk_compactness}).

Our inactive galaxy reference values for Sb spirals already include
the correction for inclination, which can be a strong source for bias:
Due to missing high inclination and edge-on spirals as observable QSO
hosts, the average field galaxy population will always appear slightly
dust-reddened in comparison. When comparing our host galaxies to the
galaxy sample used by \citet{fioc99} this amounts to bluer colours by
about 0.15~mag. In elliptical galaxies, however, this kind of dust
extinction is not expected, thus colour and viewing angle should not
be related.

After assessing all of these potential error sources, we conclude that
the blue colours of our sample -- approximately normal for the discs,
abnormally blue for the ellipticals -- are most likely real.

\subsection{Colour-magnitude relation for host galaxies}
\begin{figure}
\begin{center}
\includegraphics[bb = 40 52 313 314,clip,angle=0,width=6.9cm]{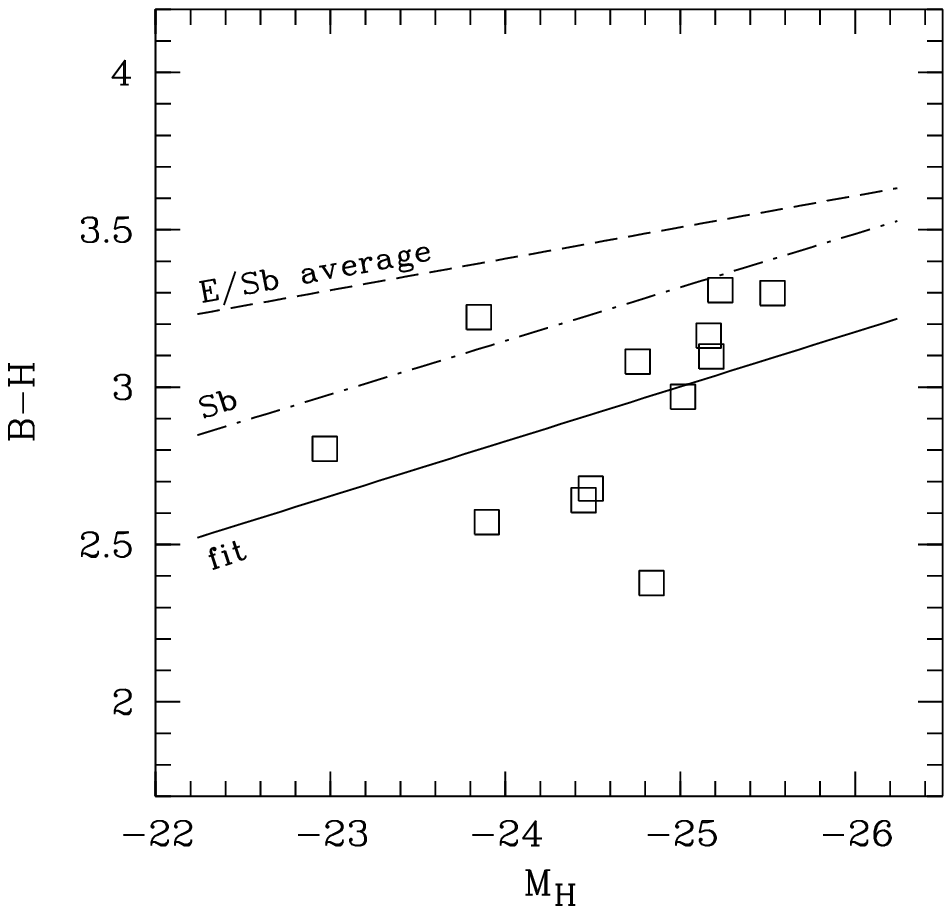}
\includegraphics[bb = 40 52 313 314,clip,angle=0,width=6.9cm]{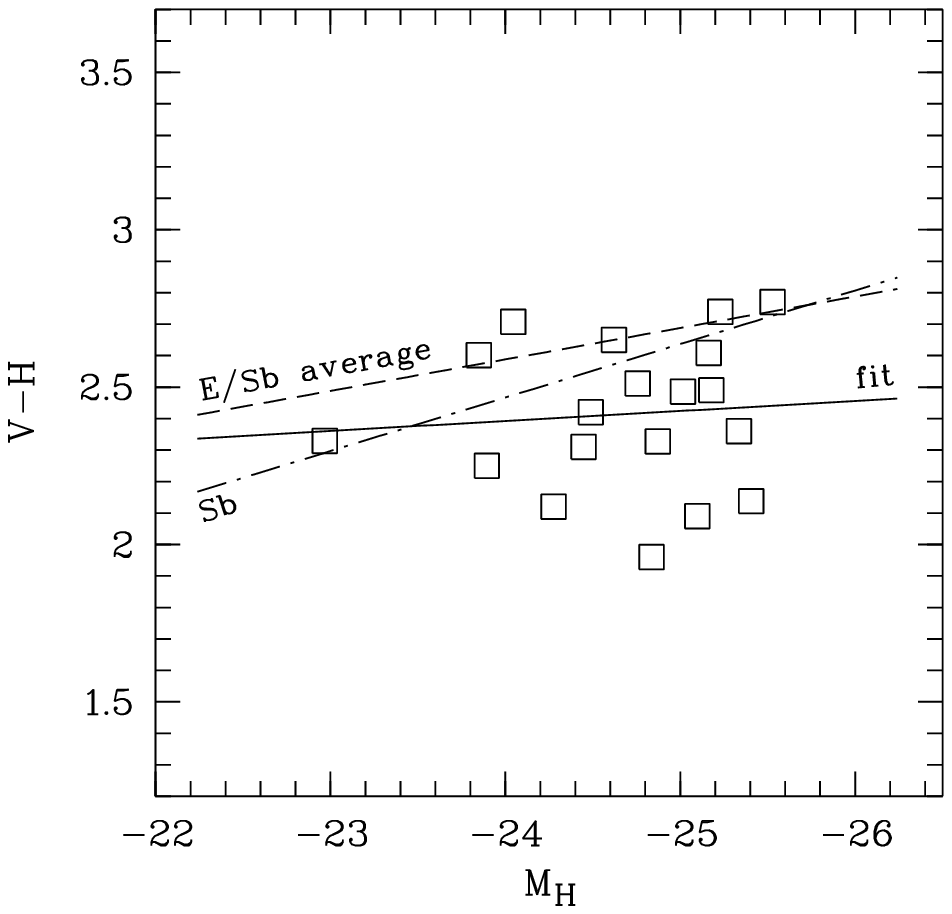}
\includegraphics[bb = 40 52 313 314,clip,angle=0,width=6.9cm]{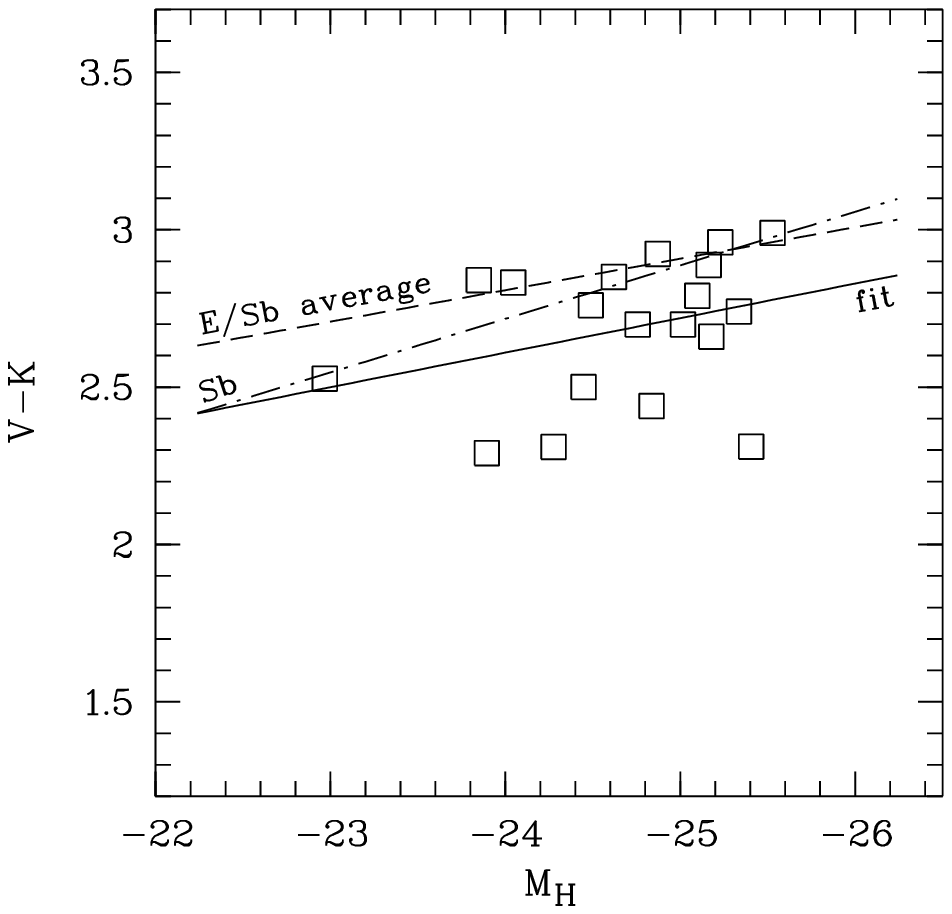}
\caption{\label{fig:col_mag} 
Colour--magnitude relations for the sample. As in
Fig.~\ref{fig:vh_disk_ellipse} optical--NIR colours against $H$-band
absolute magnitude $M_H$ are plotted for the individual objects in the
sample, now in the $V$ related colours a 0.15~mag correction is
applied for line emission in the ellipticals. Top is $B-H$ (12
objects), middle $V-H$, bottom $V-K$. The solid line is the linear
regression fit of colour and absolute magnitude of the sample. The
dashed and dot-dashed lines are the $B-H$ vs.\ $M_H$ colour--magnitude
relations for inactive galaxies (E and Sb average and pure Sb
respectively) as derived by \citet{fioc99}. In the second and third
panel these lines are shifted to represent $V-H$ and $V-K$.
}
\end{center}
\end{figure}

One further interesting aspect is shown in Fig.~\ref{fig:col_mag},
where we plot the optical--NIR colour-magnitude relation (CMR) for
three colours and the absolute $H$-band magnitude. Compared to
Fig.~\ref{fig:vh_disk_ellipse} we corrected the $V$-band related
colours of the ellipticals for the abovementioned bias due to $H\beta$
by globally adding 0.15~mag. Shown are the linear regression fits to
the colours with $M_H$ as an independent variable (solid line). This
we compare to the CMR of inactive Sb galaxies (dot-dashed line) and an
average CMR of E and Sb galaxies, both taken from Fioc \&
Rocca-Volmerange. For the latter we averaged their slopes for E and Sb
galaxies and shifted the relation to a zeropoint according to the mean
E and Sb $B-H$ colour, as given in Table~\ref{cols_sample}. For the
CMR in $V-H$ and $V-K$ we applied offsets of 0.82~mag ($B-V$ average
for E and Sb) and 0.68~mag for the Sb relation. Accordingly, offsets
of 0.60 and 0.43~mag were applied for $(B-V)-(H-K)$. While the
correlations are not strong, the similarity between the CMRs for hosts
and inactive galaxies are appealing, particularly in $V-K$, where the
slope of the fit is identical to the averaged inactive CMR relation,
just shifted towards bluer colours by 0.2~mag. It should be noted that
also the spread of the CMR of our quasar hosts is within 15 per cent to the
spread found by Fioc \& Rocca-Volmerange.

\section{Fitting stellar populations models}
\label{mc_stelpop}
Another way of analysing the derived host galaxy luminosities and
colours is by comparing them to synthetic model SEDs. In order to
combine all colour information we decided to fit SEDs of evolution
synthesis models (ESM) to the calibrated fluxes of the hosts at the
different wavelengths. With seven bands representing seven data points
in the SED of an object, our prime goal was to reproduce and
cross-check the blue colours found, and to try to estimate approximate
ages of the dominant populations in the hosts. For this task we used
model single stellar populations (SSPs) of single metallicity, because
the age-metallicity degeneracy in model colours cannot be resolved
with multicolour data. We made the simplifying assumption that the
populations are created instantly at age zero with an initial mass
function and then evolve passively (`instantanous burst
models'). Even if the absolute calibration of these model fits is
difficult, they can show if there is or is not a generally different
SED found for the elliptical and disc dominated host galaxies.

\subsection{Evolution synthesis models}\label{cols_modelpredictions}
Synthesised galaxy SEDs composed from libraries of stellar spectra can
predict colours, as a function of age and metallicity. Different
methods are used in evolutionary population synthesis and
discrepancies arise particularly in the NIR. \citet{mara98} compared
models by \citet{bruz93,bruz96} with own models and explained a
0.5~mag difference in $V-K$ colours between the two models by the
different treatment of AGB and carbon stars in the models. For single
stellar populations of about 1~Gyr, Maraston's models are redder by
several tenths of a magnitude in $V-K$ colours. For models older than
4~Gyrs the predicted colours converge again, with differences less
than 0.2~mag between the models. For SSPs older than 8~Gyrs the
numbers are also consistent with colours derived for evolved
ellipticals by \citet{fioc99}. We note that the different existing
stellar population models have not yet finally converged; however, for
a qualitative SED analysis the current state seems acceptable.

We chose to use the models from the GISSEL96 library
\citep{bruz93,bruz96}, also available via \citet{leit96}, with the
\citet{scal86} initial mass function and solar metallicities for
reasons of comparability with \citet{mara98}. We selected a coarse
grid of model ages, 0.1, 0.7, 2, 6 and 14~Gyr, and in addition
constructed a rough continuous star formation (CSF) model with a
constant star formation rate, by superposing a fine grid of models
between ages of 10~Myr and 14~Gyr.

For these model spectra broad band fluxes were computed for the seven
{\it BVRIJHK} filters used, and we converted the host galaxy
magnitudes to flux densities using a Kurucz Vega SED as given by
\citet{bruz96}. Then models and data were compared via $\chi^2$
minimisation, in two steps: 1) fitting only one SSP, with age as the
only free parameter (one-component, `1c'), 2) fitting two SSPs, with
the first age fixed (0.1 Gyr), and the age of the second component and
the two masses as free parameters (two-component, `2c').

\subsection{Variances and confidence intervals}
In the fitting process we need a weighting scheme for the
contributions of the different bands. Also, only knowing the age of
the best fit SSP is of limited diagnostic value without at least a
rough estimate of a range of ages also compatible with the data set.

Statistically correct variances are not available for this purpose, as
they are lost in the QSO deblending procedure. We attempted to
estimate the absolute values of the standard deviation $\sigma$ from
the $\chi^2$ minimisation itself. For this we followed \citet{pres95}
by not using the determined $\chi^2$ as a goodness of fit
parameter. Instead we determined the range of potential models by
using the fact that for `moderately good' fits the $\chi^2$ value is
equal to the number of degrees of freedom $\nu$, yielding a known
relation between $\sigma$, $\nu$ and $\Delta \chi^2$.

This method has the drawback of requiring identical weights for all
data points, in spite of systematically different flux values for the
seven bands. The observations were designed to yield similar S/N for
the host in all bands, except $B$, thus the \emph{relative} errors
from the deblending process are similar. This makes the
\emph{absolute} errors in the lower flux NIR bands smaller, but we
assign the same absolute error with this weighting scheme. As a result
the NIR bands will have a systematically lower weight. This effect is
counteracted by the redundancy of information contained in the three
NIR bands, all of which contain information about the older stellar
population.

\subsection{One component fits}

We performed two different 1c fits. The first used only the five SSP
of ages 0.1, 0.7, 2, 6 and 14~Gyr, in the second we added the CSF
model. In Table~\ref{tab:mc_pop1fit} we show the result: We marked the
best fitting model with an `x', those within $1\,\sigma$, $2\,\sigma$,
and $3\,\sigma$ according to the above procedure with 1, 2, and 3. As
described above, these error bars should not be taken at face value,
but only give a rough impression on the range of permitted ages.

\begin{table*}
\caption{\label{tab:mc_pop1fit}
Single SSP (1c) fit. `x' marks the best fitting profile, 1, 2, and 3 the
profiles that are acceptable within $1\,\sigma$, $3\,\sigma$, and
$3\,\sigma$ error accordingly. Also shown again is the galaxy type
(E)lliptical or (D)isk from Tab.~\ref{tab:objects2} for
comparison. The last coloumn displays a `y' if a CSF model, when
incorporated into the fit, would be accepted as best fitting model.
}
\begin{center}
\begin{tabular}{lccccccc}
Object 		&Type	&0.1~Gyr&0.7~Gyr&  2~Gyr&  6~Gyr& 14~Gyr&CSF\\
\hline          
HE\,0952--1552 &D	&	& 	&x	&3	& 	&y\\
HE\,1019--1414 &D	&	& 	&x	&3	& 	&\\
HE\,1020--1022 &E	&	& 	&1	&x	&3	&\\
HE\,1029--1401 &E	&	&x	&1	& 	& 	&\\
HE\,1043--1346 &D	&	&3	&x	& 	& 	&y\\
HE\,1110--1910 &E	&	&1	&x	&1	& 	&\\
HE\,1201--2409 &E	&	&x	&2	& 	& 	&\\
HE\,1228--1637 &E	&	&x	&1	& 	& 	&\\
HE\,1237--2252 &D	&	&3	&x	& 	& 	&y\\
HE\,1239--2426 &D	&	&1	&x	&3	& 	&y\\
HE\,1254--0934 &D	&	&2	&x	&1	&3	&\\
HE\,1300--1325 &E	&	& 	&x	& 	& 	&\\
HE\,1310--1051 &D	&	&x	&1	& 	& 	&\\
HE\,1315--1028 &D	&	& 	&1	&x	&3	&\\
HE\,1335--0847 &E	&	&1	&x	& 	& 	&\\
HE\,1338--1423 &D	&	&2	&x	& 	& 	&y\\
HE\,1405--1545 &D 	&	&x	&1	& 	& 	&\\
HE\,1416--1256 &E	&	&1	&x	&2	& 	&\\
HE\,1434--1600 &E	&	& 	&x	&3	& 	&\\
\end{tabular}  
\end{center}
\end{table*}

It is interesting that the best fitting models are generally quite
young. Only for two objects the best fit is older than the 2~Gyr
model. One of them, HE\,1315--1028, is of disc morphology, the other,
HE\,1020--1022, is an elliptical. Of the rest, twelve prefer the 2~Gyr
model, five even the 0.7~Gyr model, whereof three are ellipticals, two
are discs. The data for HE\,1254--0934 seem to constrain the models
only poorly, as all except the youngest model are possible within
$3\,\sigma$.

In a second step we therefore included a CSF model in the 1c fit,
assuming a constant star formation rate over a period of 14~Gyr. The
last column `CSF' in Table~\ref{tab:mc_pop1fit} marks those objects
with a `y', for which the CSF is the best fitting model when
incorporated in the fit. This is the case for half of the ten discs
and none of the ellipticals. Included in these five discs are the
three with visible prominent spiral arms, HE\,1043--1346,
HE\,1239--2426, and HE\,1338--1423. For four of the five remaining
discs the CSF model ranges on rank number two, all inside
$1\,\sigma$. Similarly, for the ellipticals the CSF model is not
inside $1\,\sigma$ for only two of nine objects, HE\,1029--1401 and
HE\,1201--2409.

\subsection{Two component fits}
\begin{figure*}
\begin{center}
\includegraphics[bb = 140 85 511
242,clip,angle=0,width=12cm]{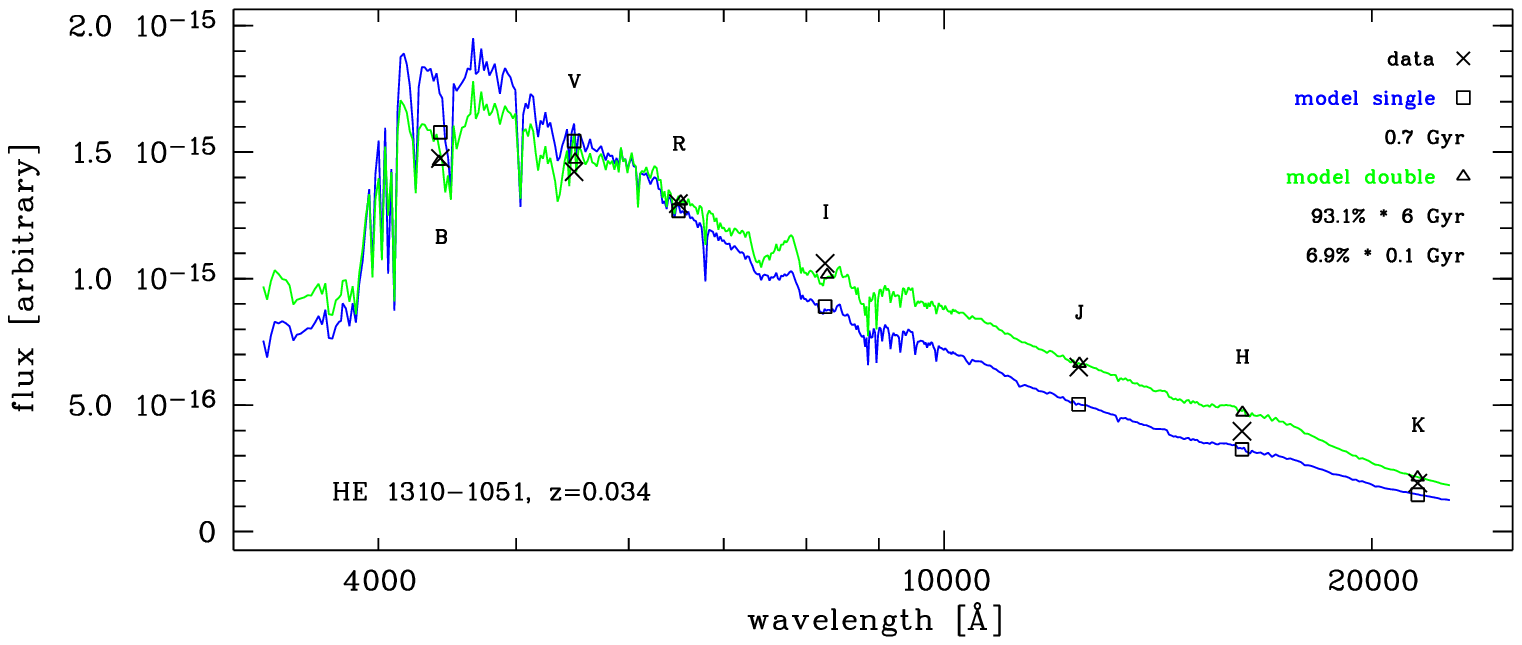}
\includegraphics[bb = 140 60 511
242,clip,angle=0,width=12cm]{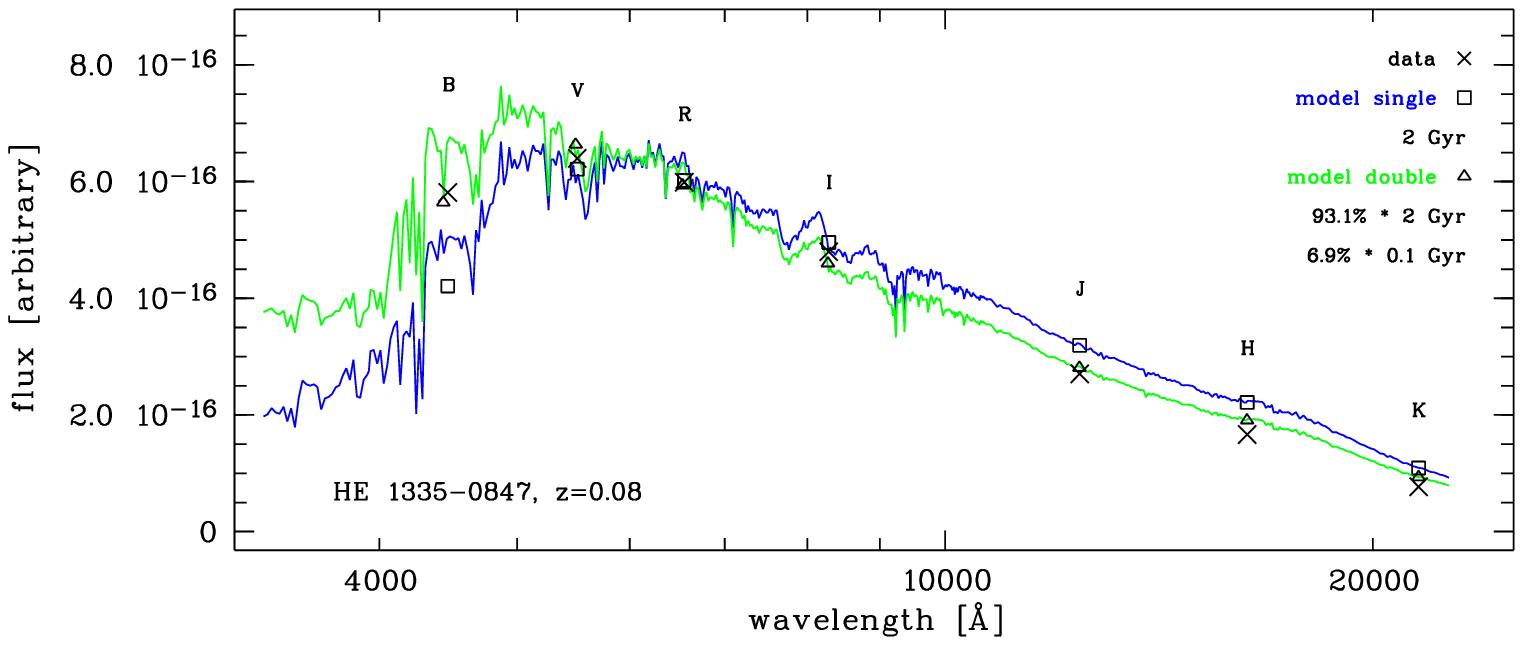}
\end{center}
\caption{\label{fig:specexample}
Two examples for results from SSP fitting. Plotted are flux densities
versus wavelength in the observed frame. Data points of the {\it
BVRIJHK} bands are marked with crosses. The SED of the best fitting 1c
SSP is plotted in dark/blue, the composite 2c SED from a young,
0.1~Gyr SSP and an older population in light/green. The broad band
averaged fluxes of the SEDs are marked with symbols: squares for 1c,
triangles for 2c. Name of the object and redshift $z$ are given, as
well as the ages and relative contributions of the SSPs.
}
\end{figure*}

For the 2c fit we proceeded similarly, with a first fit excluding the
CSF model. The age of the first component was set to 0.1~Gyr. The age
of the second component was left free, as well as both absolute scale
parameters (`masses') that we translate into a percentage of the total
(visible baryonic) mass. Results are compiled in
Table~\ref{tab:mc_pop2fit}, in a similar way to the 1c fit, with the
mass fraction of the 0.1~Gyr component added in parentheses. As
examples, the resulting best fitting SEDs and their relation to data
points are plotted for two objects in Figure~\ref{fig:specexample}.

\begin{table*}
\caption{\label{tab:mc_pop2fit}
Two component SSP (2c) fit, one component fixed to
0.1~Gyr. Nomenclature as in Table~\ref{tab:mc_pop1fit}, only added in
parentheses is the fraction of the total mass of the 0.1~Gyr component
in percent. In the CSF column, we now also list the confidence level
and percentage, even if the CSF model was not the best fit.
}
\begin{center}
\begin{tabular}{lclllll}
Object 		&Type	&0.7~Gyr&  2~Gyr&  6~Gyr& 14~Gyr&CSF\\
\hline          	
HE\,0952--1552 &D	& 	&2 (0.7)&1 (3.3)&x (2.3)&1 (0.6)\\
HE\,1019--1414 &D	& 	&x (0.8)&1 (2.6)&1 (3.3)&1 (0.0)\\
HE\,1020--1022 &E	& 	&1 (0.0)&x (1.0)&2 (1.5)&2 (0.0)\\
HE\,1029--1401 &E	&1 (0.0)&x (10.7)&1 (11.3)&1 (8.7)&1 (7.5)\\
HE\,1043--1346 &D	&	&1 (3.9)&x (5.2)&2 (3.3)&y (1.7)\\
HE\,1110--1910 &E	&2 (0.0)&x (2.6)&1 (4.8)&1 (4.5)&1 (2.3)\\
HE\,1201--2409 &E	&x (0.0)&1 (12.3)&1 (10.7)&2 (7.5)&1 (8.7)\\
HE\,1228--1637 &E	&	&x (5.6)&1 (6.5)&2 (4.5)&1 (3.9)\\
HE\,1237--2252 &D	&	&2 (3.3)&x (4.8)&3 (3.1)&1 (1.3)\\
HE\,1239--2426 &D	&	&3 (6.5)&1 (5.3)&x (4.1)&1 (3.3)\\
HE\,1254--0934 &D	&2 (0.0)&x (0.7)&1 (3.6)&2 (3.1)&1 (0.0)\\
HE\,1300--1325 &E	&	&1 (0.7)&x (3.8)&	&y (1.1)\\
HE\,1310--1051 &D	&	&2 (7.0)&x (7.0)&2 (5.2)&1 (4.5)\\
HE\,1315--1028 &D	&	&2 (0.0)&x (1.1)&2 (2.3)&2 (0.0)\\
HE\,1335--0847 &E	&	&x (7.0)&2 (8.7)&3 (6.0)&2 (5.2)\\
HE\,1338--1423 &D	&	&1 (2.4)&x (4.5)&1 (3.3)&1 (1.8)\\
HE\,1405--1545 &D 	&1 (0.2)&x (9.3)&1 (8.7)&2 (7.0)&1 (6.5)\\
HE\,1416--1256 &E	&1 (0.2)&x (5.2)&1 (7.5)&1 (6.0)&1 (5.2)\\
HE\,1434--1600 &E	& 	&x (0.7)&1 (4.5)&1 (3.9)&1 (0.9)\\
\end{tabular}  
\end{center}
\end{table*}

\subsection{SSP fitting results}\label{sec:popfitresults}
In total the objects show broad band colours consistent with
intermediately young stellar populations, with an age of $\sim$2\,Gyr
or CSF. It is again striking that hosts of both morphological types
are consistently fit by relatively young spectra. Significant
contributions of a very young population is required by only two discs
and no significant difference can be seen for spheroids vs.\ discs
concerning average age. {\em None} of the spheroids shows an SED that
requires to be modelled with an old, evolved population, as would be
expected for inactive ellipticals.

This result is supporting the blue $V-K$ colours found in
Section~\ref{mc_colourcomparison}. Colours as blue as seen for most
objects are usually only found in late type inactive galaxies that
have a significant component of ongoing star formation. For the discs
in the sample we should expect most of the blue emission to come from
continuous star formation as normal for spirals, but for the spheroids
this is not expected. The on average slightly bluer $V-H$ colours for
spheroids compared to discs are compatible with the result of the
model fitting. Also, the ages of the dominating populations in the
spheroids are compatible with the abnormously blue colours found.

\section{Discussion}\label{sec:astrodiscussion}

\subsection{Host luminosities and morphological types}
Our sample is statistically complete, thus representative for the
local QSO and QSO host galaxy population. We do not sample the highest
luminosity regime for both components; in $V$-band luminosities the
sample is distributed around the classical dividing line between Sy\,1
galaxies and luminous quasars.

In our sample about half of the objects are ellipticals, the other
half is disc dominated. For this range of quasar luminosities, the
morphological classification found for the sample agrees well with
previous findings \citep[e.g.][]{smit86,mcle95a,tayl96,scha00}: The
lower the luminosity of the nucleus, the larger the probability to
find a disc as the host galaxy; the higher the quasar luminosity, the
more probable it becomes to find an elliptical. In absolute
luminosities of the hosts (Table~\ref{objects3}), the ellipticals in
our sample are brighter than the discs, on average by 0.4~mag in $V$,
0.2~mag in $K$.

\begin{figure}
\begin{center}
\includegraphics[bb = 71 572 349 772,clip,angle=0,width=\colwidth]{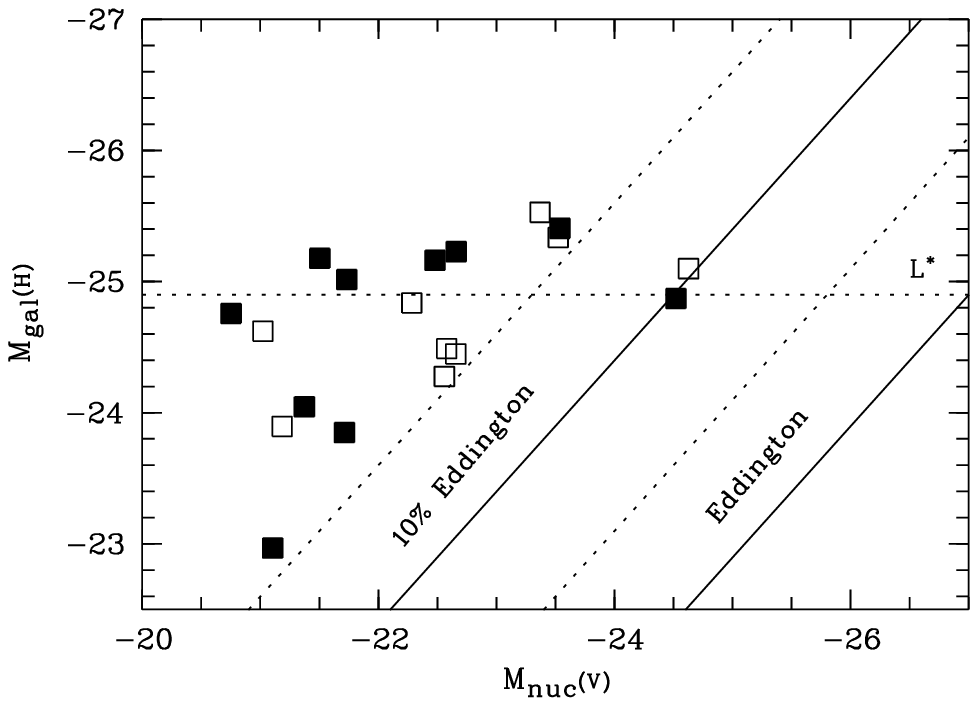}
\end{center}
\caption{\label{mnuc_mgal}
Host $H$-band magnitude vs. nuclear $V$-band magnitude for the
multicolour sample. Open symbols mark ellipticals, filled symbols
discs. The dotted line shows the magnitude of an $L^*$ galaxy in the
$H$-band, the solid diagonal lines mark the loci of Eddington (right)
and 10 per cent Eddington (left) luminosities for galaxies following the
\citet{mago98} relation $M_\mathrm{bh}=0.006\cdot M_\mathrm{sph}$
\citep{mcle99}. If more recent values are used for the relation
between black hole and bulge luminosity, $M_\mathrm{bh}=0.002\cdot
M_\mathrm{sph}$ \citep[see][and referenced therein]{dunl03}, then the
dashed diagonal lines apply.
}
\end{figure}

In the $H$-band about half of our objects are brighter than $L^*$ for
inactive galaxies \citep[$K$-band values taken from][and assuming $H-K
= 0.25$~mag]{huan03}, as shown in Figure~\ref{mnuc_mgal}, thus are
generally luminous. Via the Magorrian (\citeyear{mago98}) relation
between the mass of the central black hole and the bulge mass of the
galaxy we also expect to find on average brighter hosts for brighter
nuclei \citep{tayl96,mcle99}. In Figure~\ref{mnuc_mgal} we find this
relation confirmed by our sample, showing a correlation of host and
nuclear luminosities. All of our hosts accrete with less or equal to
10 per cent of their Eddington luminosity, independent of morphology
and host luminosity.

As noted in Sect.~\ref{sec:mcresultmorph}, we lack discs with
ellipticities of $<0.1$ and $>0.5$. Similar results were found in
several other host galaxy studies. \citet{tayl96} find only one out of
eleven discs with $e<0.1$ and a highest ellipticity of
$e=0.51$. \citet{mcle95b} investigate a large Sy~1 sample, and also
find a strong bias against edge-on galaxies, with only four out of 42
Sy~1 showing $e>0.5$. All samples, including our own, peak around
$e=0.3-0.4$, which, apart from confirming the standard model
\citep{anto93}, suggests an intrinsic ellipticity for the
discs.

Our distribution of axial ratios ($q=b/a=1-e$) for the ellipticals, as
shown in Fig.~\ref{fig:ellipticities}, is very similar to numbers
found by \citet{trem96} \citep[a similar definition was proposed
by][]{fabe97}. The authors performed statistical tests on axial ratios
on a large sample of inactive elliptical galaxies, finding two
distinct populations of ellipticals, dependent on luminosity. They
separate between faint and bright ellipticals at $M_B\sim-20.5$, which
puts all of our elliptical hosts into the latter class. The mean axial
ratios are around 0.75 for their faint sample and 0.85 for the bright
counterpart. Our ellipticals have $q=0.82\pm0.04$ when assuming
$e=0.3$ for HE~1300--1325 (see Sect.~\ref{sec:mcresultmorph} above),
with a spread of 0.11. Tremblay \& Merritt find a spread of 0.08. Thus
we find very similar values to the axial ratio distribution of bright
inactive ellipticals. From their morphological parameters alone, there
is no evidence that our hosts deviate from the inactive galaxy
population except for the lack of edge-on discs.

\subsection{Blue host colours and young stellar ages}
As we have shown in Section~\ref{sec:comp_inactive}, particularly the
elliptical host galaxies show unusually blue colours, both for short
and long wavelength baselines. In $V-K$ we see the discs being
slightly (0.14~mag) bluer than inactive Sb galaxies, but only with
marginally significance. The elliptical hosts, on the other hand, are
bluer by 0.5~mag than their inactive counterparts, even \emph{bluer}
than the discs by 0.25~mag. If we include a 0.15~mag correction for
possible line emission, the ellipticals are still bluer than inactive
ellipticals by 0.35~mag and only slightly bluer than their disc
counterparts. The colours of both morphologically defined subgroups
are similar to those of intermediate to late type spirals.

If this is correct the assumed K-corrections for the ellipticals are
overestimated. If instead the K-corrections appropriate for later type
galaxies from Table~\ref{t_kcorr} are used, the ellipticals become
even more luminous in the rest frame $V$-band. Their average redshift
is $\overline{z}=0.11$, resulting in $\sim$0.15~mag higher
luminosities and thus even bluer $V-K$ colours. Incidentally, in the
$V$-band the effects of gas emission and K-correction would cancel
out.

The results from fitting the broad band fluxes with models points into
the same direction. When including corrections for the $V$ and
$I$-band fluxes, the derived ages become slightly older, but the
results do not change fundamentally.

For several of the discs as well as some of the ellipticals a
continuous star formation model is also adequate. But \textit{none of
the ellipticals} is modelled best by an old, evolved population. On
the other hand there is little evidence for massive on-going
starbursts, manifested in a significant young population ($\ll
1$~Gyr). Only for two objects starburst masses above a few percent are
found -- both being discs.

A strong starburst on top of an evolved old population can be excluded
as the reason for the blue colours from the model
deblending. Interestingly, the blue colours are visible, even though
asymmetric features typically showing increased star formation have
been masked out in the deblending and photometry process. In fact,
there could be even stronger star forming activity going on in those
regions.

\subsection{Comparison to other studies}
The number of existing host galaxy studies featuring any sort of
spectral information is small. Only rarely, more than two bands (one
colour) is available. Results concerning evidence for young stellar
populations or starbursts are mixed and show that stellar age and
morphological classification are not linked in a simple
manner.

\begin{itemize}
\item The lower luminosity regime of classical Sy\,1 galaxies was
investigated by \citet{koti94}. They observed a sample in the
\emph{BVRIJHK} bands, similar to us. As expected for Seyferts, they
found only spiral galaxies. All of their objects show very normal
colours for their morphological type and no signs for additional
starbursts. This is in agreement with our results for disc galaxies
when continued towards lower luminosities.

\item \citet{scha00} investigated a large sample of 76 low redshift
($z\le0.15$) predominantly intermediate luminosity AGN. They observed
from the ground in the $B$ and $R$-bands and with {\it HST} in the
F814W ($\sim$$I$) band. Their objects show host galaxy properties
similar to our the lower luminosity objects in our sample. No strong
merger events are observed, only central bars in a number of
objects. Individual $B-I$ colours of their host galaxies show a large
spread, which in part is stated to be due to the deblending of nucleus
and host galaxy components. Without going into detail -- this topic
not being their primary focus --, the authors report for their sample
morphological types and colours not different from inactive galaxies.
However, when we calculate the distribution of their $B-I$ colours,
differentiated by morphological type, we find E/S0 colours to be
$\sim0.4$~mag bluer than inactive (Fig.~\ref{schade}). Even though a
large colour spread is found, both the mean value as well as the
majority of individual objects lie bluewards of inactive galaxies.

\begin{figure}
\begin{center}
\includegraphics[bb = 73 451 537 772,clip,angle=0,width=\colwidth]{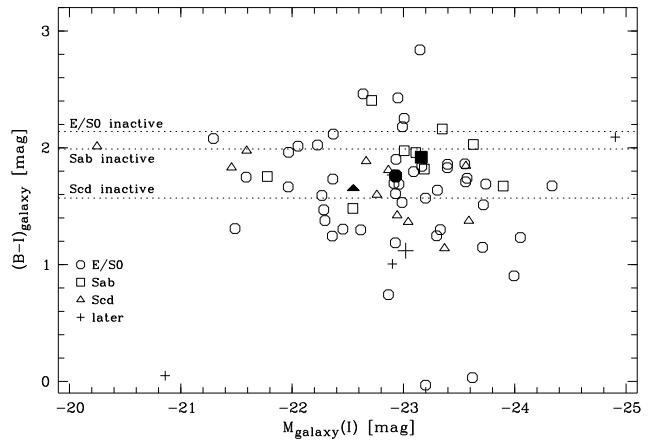}
\end{center}
\caption{\label{schade}
Rest frame $B-I$ host galaxy colours from \citet{scha00}, recomputed
from their given apparent magnitudes, converted to Vega zeropoint,
K-corrected and differentiated by galaxy type according to bulge/total
ratio. Individual objects are marked by open symbols, mean type values
by larger filled symbols and cross. The horizontal lines are the mean
inactive galaxy colours from \citet{fuku95}.
}
\end{figure}

\item In a large $B$-band study with 53 objects \citep{jahn03e}, we
compared host galaxy luminosities to the NIR sample properties of
\citet{mcle94a,mcle94b}. The sample in that study includes the present
multicolour objects. We found indications for $\sim$1~mag bluer $B-H$
colours than expected for inactive galaxies in the
$M_{B,\mathrm{tot}}<-23$ high luminosity subsample. The lower
luminosity galaxies were consistent with inactive galaxy colours.

\item \citet{roen96} and \citet{oern03} investigated a large sample of
intermediate redshift quasars ($0.4<z<0.8$), for which already some
evolution might be expected. They also attempted to determine
morphological information and found a few elliptical RQQ host galaxies
with similar luminosities than our luminous objects, and rest frame
$B-V$ colours bluer than expected by 0.6~mag, much more consistent
with late type spiral (Sc--Sd) colours.

\item The first thorough spectroscopic study of QSO host galaxies
\citep{boro85} observed a heterogenous sample of quasars from
intermediate to high luminosites using off-nuclear spectroscopy. There
was no morphological information available to classify the hosts. They
identified different classes of host galaxies, with either blue or red
continuum, pointing to both young and old populations, respectively. A
comparison to the spectra and luminositites of \citet{nola01} suggests
that their class of `red-continuum' objects can be identified as
elliptical galaxies with old evolved stellar populations. However
there exists another class of `blue-continuum' objects, whose
morphological identity is not known.

\item In a program started ten years ago
\citep{dunl93,tayl96,mclu99,dunl03}, the spectroscopic study by
\citet{hugh00} and \citet{nola01} investigated each nine radio-loud
and -quiet quasars at redshifts $z\la0.3$. These quasars are on
average 1.5~mag more luminous compared to our sample, although an
overlap in luminosities exists. They found elliptical host galaxies
being more luminous in the NIR by 1.2~mag \citep{tayl96,dunl03}
compared to our objects, with scale lengths being larger by a factor
of two \citep{dunl03}. With off-nuclear spectra they sampled the host
galaxies at relatively large distances to the nuclei of 15--20~kpc,
attempting to avoid the majority of nuclear emission in their
spectra. Their extracted host galaxy spectra are characteristic for
intermediately old to old populations, their youngest estimated
stellar ages being several Gyrs. They find little gas emission and no
signs for young populations or starbursts above $\sim 1$ per cent in
mass.  $R-K$ broad band colours of a subsample of these objects are
presented in \citet{dunl03}. They also generally the hosts to be red
and conclude that old populations dominate.

Incidentally, one of their faintest and our brightest elliptical host
is identical. For this object, HE\,1020--1022, both $K$ and $R$-band
magnitudes are consistent, with a resulting colour difference of only
$\Delta(R-K)=0.05$~mag. For precisely this object \citeauthor{nola01}
found the youngest population in their sample. Their best fits
suggested either a 5~Gyr component plus 0.6 per cent of 0.1~Gyr, or a pure
4~Gyr population when a contribution for the nucleus is
subtracted. With multicolour fitting we found 6~Gyr plus 1 per cent of
0.1~Gyr and compatibility with a pure 2~Gyr component. The 6~Gyr fit
would make this object the one with the oldest dominating population
in our sample, and, independent of this, it shows also the reddest
$V-K$ colour.

\end{itemize}

The morphology and colours of the disc host galaxy fraction of our
sample seem to be in general good aggrement with these previous
results as summarized above. The lower luminosity Sy~1 sample of
\citet{koti94}, the discs from \citet{scha00} as well as the discs
found by \citet{roen96} at intermediate redshifts show largely normal
colours compared to their inactive counterparts, as do our discs. It
is difficult to assess whether this still holds for disc host galaxies
of higher luminosities, due to their rapidly decreasing numbers above
$L^*$.

On the other hand, the colours of the ellipticals in a certain
luminosity range seem to differ from their inactive counterparts. Both
the ellipticals from \citeauthor{scha00} as well as in this study
appear bluer. However, at higher luminosities the elliptical hosts, as
investigated by \citet{dunl03}, show perfectly inactive colours. Thus
at high luminosities there exists at least one population of
apparently normal elliptical host galaxies that have not recently
(0.5~Gyrs) been involved in strong star formation.

\subsection{Is it interaction?}
Several known quasars show obvious signs of on-going merging with a
second galaxy. 3C~48, one of the first quasars identified, has been
thoroughly investigated with long-slit and integral field spectroscopy
\citep{cana00a,chat99}. It was found that it is involved in a late
stage of a merger, showing conversion of gas into new stars in local
starbursts in several regions, turbulent flows of gas and a secondary
nucleus. The starburst regions in the host of 3C\,48 and similar
objects show very blue colours. In our sample we find apparent signs
of interaction like tails, asymmetries or double nuclei in a
substantial fraction ($\sim$50 per cent, 9/19) of our objects. Strong violent
merger configurations are only apparent in a few of these, such as
HE~1405--1545 or HE~1254--0934, and most of them do not appear as
violent as 3C\,48. The high fraction of host galaxies in merger state
supports the picture that interaction might be important for fuelling
of the nucleus. However, the existence of very symmetric and
undisturbed host galaxies like HE\,1043--1346 (spiral) and
HE\,1029--1401 (elliptical) shows that ongoing major mergers are not a
prerequisite for quasar activity. Yet the observed blue colours of
especially the elliptical hosts suggest that some unusual event in the
not so distant past may be connected to the onset of nuclear activity.

This either requires a significant time delay (at least several
100~Myr) between the event and the ignition of the nucleus, or a very
long duty cycle of the nuclear activity, or other mechanisms than {\em
major} mergers to play a significant role. In any case, for Sa--Sc
host galaxies the trigger for the nuclear activity cannot be major
mergers, as these would destroy the observed disc structures. Thus
probably minor mergers and maybe even quiescent gas accretion play a
significant role in the nuclear fuelling.

Minor merger as the trigger could also explain qualitatively the
correlation of colour and luminosity (or mass). An interaction with a
companion of a given mass will have stronger effects on the lower mass
than on the higher mass ellipticals, resulting in larger asymmetries,
and, in the presence of gas, relatively stronger induced
star formation.

\section{Conclusions}
Are QSO host galaxies just normal galaxies with an active nucleus? We
find strong evidence that at least a subpopulation has significantly
different colour properties than their inactive counterparts. While
the disc dominated or spiral galaxies in our sample show statistically
no difference in colour and age to inactive discs, the ellipticals
exhibit abnormally blue colours and SEDs consistent not with old,
evolved stellar populations. 

Even though elliptical galaxies can be significantly more luminous
than disc galaxies, the abnormally blue colours found are not a
function of only luminosity. Both for our sample as well as that by
\citeauthor{scha00}, the two classes of bulge vs.\ disc dominated
hosts contain objects of similar luminosity (see
Figs.~\ref{fig:vh_disk_ellipse} and \ref{schade}). This suggests that
the main difference lies in the {\em morphological type} of the host
galaxy. However, at some point a luminosity -- or mass -- dependence
has to exist, since the more massive elliptical host galaxies studied
by \citeauthor{dunl03} show old evolved stellar populations and signs
of only small recent star formation.

The correspondence between the QSO activity and the blue colours
suggests a connection between galaxy interaction, induced star
formation, and the onset of nuclear activity. However, we find that it
is likely that mechanisms other than major mergers, such as minor
merger or gas accretion, are responsible for the activity of both the
high luminosity ellipticals as well as the disc-dominated host
galaxies.

At this point the picture of how activity and interaction are
connected is not clear. However, the abnormally blue colours of a
sub-population are a strong indicator that for spheroid-dominated host
galaxies, merging is an important factor in triggering nuclear
activity.

\section*{Acknowledgments}
The Nordic Optical Telescope is operated on the island of La Palma jointly
by Denmark, Finland, Iceland, Norway, and Sweden, in the Spanish
Observatorio del Roque de Los Muchachos of the Instituto de
Astrofisica de Canarias.

KJ gratefully acknowledged support by the Studienstiftung des
deutschen Volkes. This work was also partly supported by DFG grants
Wi~1369/5--1 and Re~353/45-3. Thanks go to Eva \"Orndahl for
discussions and the jointly taken NOT data, and to Martin Raue for
compiling the companion statistics.

\bibliographystyle{mn2e}
\bibliography{knuds}

\label{lastpage}

\end{document}